# Corrosion Evolution of T91 Steel in Static Lead-Bismuth Eutectic Under an Oxidising Environment

Minyi Zhang[1,2]*, Weiyue Zhou[4], Michael P. Short[4], Paul A.J. Bagot[1], Michael P. Moody[1,5], and Felix Hofmann[3]*

[1.] Department of Materials, University of Oxford, Parks Road, Oxford, OX1 3PH, United Kingdom.

[2.] Max-Planck-Institute for Sustainable Materials, Max-Planck-Straße 1, Düsseldorf, 40237, Germany

[3.] Department of Engineering Science, University of Oxford, Parks Road, Oxford, OX1 3PJ, United Kingdom.

[4.]Department of Nuclear Science and Engineering, Massachusetts Institute of Technology, 77 Massachusetts Avenue, Cambridge, MA 02139, USA.

[5.]Australia's Nuclear Science and Technology Organisation, New Illawarra Road, Sydney, Lucas Heights NSW 2234, Australia.

* Corresponding authors:

Minyi Zhang (m.zhang@mpi-susmat.de, +49 1623298641),

Felix Hofmann (felix.hofmann@eng.ox.ac.uk , +44 1865283446)

## Abstract

Understanding corrosion in liquid metal-cooled nuclear systems is essential in order be able to control it. While much literature exists detailing corrosion rates and mechanisms of structural materials in liquid metals, much still remains to be discovered in new regimes of temperature, chemistry, and impurity content. We focus on a less-studied set of conditions, specifically to investigate how liquid lead-bismuth eutectic (LBE) corrodes ferritic/martensitic steels under high-temperature oxidizing conditions. We find that the evolution of corrosion is determined by the formation of protection layer on the surface. The area without effective protection layer experiences oxidation along martensite grain boundaries, transitioning from intergranular attack to broader area corrosion as it progresses. The area a stable, coherent oxide scale will slow the corrosion process and then is oxidized along pre-austenite grain boundaries. Both chromium and oxygen diffusion play vital roles in this process. Most surprisingly, a layer of iron enriched body-centred cubic phase forms on the surface of LBE-corroded T91, contradicting previous studies, which reported only oxide-based surface layers.

## 1. Introduction

Liquid metals, such as lead–bismuth eutectic (LBE), pure lead (Pb), and lead–lithium (PbLi) alloys, are considered promising candidate coolants for Generation IV fast nuclear fission reactors. These coolants possess several advantageous properties [1], including low neutron moderation and capture cross-sections, low melting points, low vapor pressures, wide margins to boiling, excellent gamma radiation shielding, and relatively low reactivity with water and air [2]. Interest in these systems has been further stimulated by the growing global emphasis on mitigating climate change and reducing $CO_2$ emissions [3].

T91, a ferritic–martensitic (F/M) steel, is a leading candidate structural material for these high-temperature energy systems. It exhibits excellent resistance to irradiation-induced void swelling [4], along with low thermal expansion, relatively high thermal conductivity, widely tuneable strength, good creep resistance (critical for high-temperature operation), and a well-established supply chain with ASME nuclear qualification [5]. Compared to martensitic steels with higher chromium content, T91 demonstrates a smaller irradiation-induced ductile-to-brittle transition temperature (DBTT) shift. According to the research of Kohyama, at 250℃, with 0.8 dpa irradiation, DBTT shift for 9CrWVTa steels is around 24℃ while for 11CrMoVNb steels it can be as high as 175 ℃ [6]. Moreover, F/M steels such as T91 mitigate the problem of accelerated corrosion at temperatures above 500 °C, in contrast to austenitic steels where fertilization of the austenite phase is promoted [7]. In this work, we focus on the use of T91 in conjunction with LBE for LBE-cooled fast reactor (LBEFR) applications. Fast reactors are regarded as a key technology for enhancing the sustainability of nuclear energy, as they enable more efficient utilization of existing nuclear waste, allow for easier closing of the fuel cycle, and contribute to the reduction of the long-term radioactivity of spent fuel [8].

This study investigates corrosion of T91 in LBE at elevated temperatures exceeding 700 °C, which would increase electrical conversion efficiency and enable the usage of high-temperature process heat for other industrial purposes; Two important factors for the economic viability of LBEFRs. Such high operating temperatures are achievable with LBE, unlike with water, and represent one of the key advantages motivating the adoption of this coolant. However, corrosion of structural steels becomes increasingly severe under these conditions, and thus remains a major bottleneck in the further development and deployment of LBEFRs.

Passivation is a commonly adopted strategy to enhance the corrosion resistance of structural materials by forming a stable, protective oxide layer on the steel surface, thereby minimizing direct contact between the liquid metal and the substrate [9]. For example, the Russian SVBR reactor design relies on the formation of an oxide layer to protect structural components [10]. According to literature, maintaining an oxygen content above $10^{-6}$ wt. % and operating temperatures below 570 ℃ [11, 12] are typically required for the formation of a stable passivating oxide in LBE. Therefore, at the elevated temperature of 700 °C investigated in this study, continuous and protective passivation is not expected to occur for T91. This limitation incentivized previous papers [2, 13], which focused on the reducing environment with an oxygen level between the equilibrium oxygen potential of $Fe_3O_4$ and $FeCr_2O_4$ at test temperature. LBE was found to penetrate T91 without following GBs and the corrosion caused martensite grains at the surface to transform to ferrite grains. In these studies, corrosion in an oxidising environment was not studied.

Understanding the corrosion mechanisms of T91 under oxidizing conditions is crucial, since oxygen, which exists as an initial impurity in LBE and can also be introduced during operation [14], may also play a role in mitigating corrosion even at temperatures exceeding 600 °C . According to the literature, under oxygen-saturated and oxygen-controlled conditions, the surface oxide typically consists of two layers:

an outer magnetite ($Fe_3O_4$) layer and an inner Cr-containing layer, mainly composed of Fe–Cr–O spinel [15]. During the initial oxidation of fresh materials, variations in elemental diffusion and dissolution govern the formation, morphology, and stability of complex oxide structures, while the resulting oxide layers subsequently regulate further corrosion and dissolution processes [16, 17]. The growth of these layers involves both the outward diffusion of elements from the steel (for example Fe, Cr, Si) and the inward migration of oxygen ions [18]. Cracks at the oxide/substrate interface have been reported in the literature. Dionisio et al. [18, 19], attributed these cracks to differences in the thermal expansion coefficients of the substrate and oxide scale, especially after long oxidation times. Considering the cyclic operating conditions of LBE-cooled reactors, such oxide scale degradation could pose a serious concern for structural integrity.

Therefore, to improve the long-term reliability of structural materials in advanced LBE-cooled systems, investigating the corrosion behaviour of T91 under oxidising conditions (oxygen level between the equilibrium oxygen potential of PbO and FeO at the test temperature) at high temperatures is essential. In particular, we study the stability of oxide scales and aim to identify the mechanisms governing corrosion progression. While previous studies of the corrosion behaviour of steels in LBE in an oxidising environment exist, these focus predominantly on lower temperatures (<600°C) and did not fully explore behaviour across orders of magnitude in time. We present a systematic study across multiple timescales at higher temperatures, revealing the corrosion mechanisms under these conditions and the critical conditions required to provide protection through passivation.

## 2. Materials and Methods

### 2.1 Materials

T91, also known as Fe-9Cr-1Mo steel, is a ferritic/martensitic (F/M) steel with body-centred cubic (BCC) lattice structure. The materials used in this study was purchased from Edelstahl Witten-Krefeld GMBH in the quenched and tempered (Q&T) condition, having undergone heat treatment in accordance with the manufacturer's specifications [20]. The final heat treatment of T91 is normalizing + high temperature tempering, the normalizing temperature is 1040°C, the holding time is not less than 10 minutes, the tempering temperature is 730-780°C, the holding time is not less than 1h. The structure after the final heat treatment is tempered martensite. This process ensures the desired mechanical properties, including a Brinell hardness in the range 190–248 HBW, making the alloy suitable for high-temperature structural applications. No further heat treatment was carried out before the corrosion tests.

All samples were cut from the same bar using wire electrical discharge machining (EDM) to dimensions of 25 mm x 25 mm x 3-4 mm. The composition from the alloy certificate is listed in Table 1.

*Table 1: Composition of T91 material according to the alloy certificate provided by the manufacturer (wt. %).*

| Fe | Cr | Mo | Si | Mn | Ni | V | C | Nb | N | P | Al |
|---|---|---|---|---|---|---|---|---|---|---|---|
| 89.0 | 8.76 | 0.93 | 0.34 | 0.50 | 0.30 | 0.20 | 0.09 | 0.07 | 0.05 | 0.02 | 0.01 |

The lead and bismuth used in this study were obtained from Surepure Chemetals, Inc.. The original materials were ultra-high-purity lead (99.999 wt. %) and bismuth, which were then combined in accordance with the eutectic composition of lead (44.5 wt. %) and bismuth (55.5 wt. %).

**2.2 Corrosion tests**

The samples were ground with successive SiC papers (grit 120, 240, 400, 800, 1600) with water lubricant, followed by polishing with Buehler MetaDi diamond suspensions (9µm, 3µm, 1µm) adding three squirts of suspension every 20-30 seconds. Final polishing with 50 nm alumina suspension achieved a mirror finish on one side, while the other side was only ground with 120-grit SiC paper. The polishing wheel was run in the contra direction (samples and platen rotating in opposite directions) at 120 revolutions per minute (RPM), with 6 lbs. force per sample. Samples were then immersed in liquid LBE for 70, 245, or 506 hours at 700 °C. The oxygen level was maintained between the equilibrium oxygen potential of Fe oxides and Pb/Bi oxides at the testing temperature [28]. Since the oxygen potential of the cover gases was below the detection limit of the sensors used (0.1 ppm), the measured concentrations of hydrogen and water vapour were used to estimate the oxygen partial pressure and concentration. A mass flow controller was used to inject precise amounts of hydrogen, based on the far more sensitive moisture meter (1 ppb). Adding water vapor increased the oxygen concentration to provide an “oxidizing” atmosphere, while leaving water vapor out kept a “reducing” atmosphere inside the furnace. A detailed description of the static corrosion experiments is provided elsewhere [21, 22].

After exposure, the samples were cut into four pieces, producing fresh, uncorroded surfaces. The pieces were mounted in Bakelite such that the fresh surfaces faced outward, while the LBE-corroded surfaces were embedded in the resin. The samples were subsequently ground and polished to reveal cross-sections comprising the uncorroded interior and the corroded surface layers. The cross-sections were then ground using the same procedure as for initial coupon preparation above, with the exception of using colloidal silica for final polishing instead of colloidal alumina. This was done in order to provide a small degree of chemomechanical polishing and highlight grain contrast in optical and electron microscopy.

### 2.3 Characterisation methods

The morphology and chemical composition of the corroded sample cross-sections were analysed using a Zeiss Merlin FEG-SEM equipped with a Bruker XFlash FlatQUAD 5060 F EDX detector and a Bruker e-flash high-resolution EBSD detector. EDX measurements were conducted at a working distance of 18 mm with an accelerating voltage ranging from 3 to 20 kV. Lowering the accelerating voltage increases both lateral and depth resolution in EDX mapping [23], which is particularly important for detecting oxygen ($K_{\alpha} = 0.523\ keV$), whose signal is prone to edge effects at higher voltages. Overlap with Cr ($L_{\alpha} = 0.573\ keV$) can occur at lower accelerating voltage, but this is addressed through peak deconvolution. Electron Backscatter Diffraction (EBSD) maps were acquired at a working distance of 18 mm, a sample tilt of 70°, a beam voltage of 20 kV, and a probe current of 5–10 nA. The data was processed using the Esprit 2.3 software.

Additional EDX analyses were performed using a dual-beam Zeiss Crossbeam 540 FIB-SEM equipped with an Oxford Instruments XMax$^{N}$ 150 EDX detector and a Nordlys Max EBSD system. Measurements were conducted at a working distance of 5 mm, with an accelerating voltage of 5–30 kV and a probe current of 2 nA. The data were processed using the Aztec 5.1 software. For verification, higher accelerating voltages were employed to resolve potential overlaps, such as those between Mo ($L_{\alpha} = 2.293\ keV$) and Pb ($M = 2.342\ keV$) peaks.

## 3. Results

### 3.1 Differences in observed corrosion patterns

Corrosion is a complex process influenced by several microstructural factors, including surface condition, grain orientation, and grain boundary distribution among other factors [24]. Its manifestation can vary with exposure duration (70 h, 245 h, and 506 h in this study) as well as spatially within a single specimen. Fig. 1 shows representative corrosion morphologies observed by SEM with voltage of 5-20 kV using the secondary electron (SE) detector after different exposure times. Based on these observations, two distinct corrosion evolution pathways can be identified. The key factor governing these different behaviours appears to be the formation (or absence) of an effective protective surface layer, which critically influences the subsequent corrosion development.

The corrosion evolution in regions where no effective protective layer is shown in Fig. 1(a–c). The process initiates with intergranular attack along martensite grain boundaries (GBs), as shown in Fig. 1(a), and subsequently progresses to more extensive surface corrosion, as evident in Fig. 1(b) and Fig. 1(c). This form of corrosion will hereafter be referred to as “wider-area corrosion.”

Wider-area corrosion is likely formed through a process that initiates at grain boundaries and subsequently extends into adjacent grains, resulting in the development of broader corroded regions. The red arrows in Fig. 1(c) mark cracks that may have originated from the mismatch in thermal expansion coefficients between the substrate and the oxide scale [18]. These cracks become more frequent with increasing exposure time. From SEM imaging alone, it is difficult to distinguish whether these features result from liquid metal (LBE) penetration or the formation of larger oxide structures. To clarify their origin, EDX analysis was performed to assess elemental redistribution within these zones in section 3.3.

The corrosion evolution in regions where an effective protective layer has formed is presented in Fig. 1(d–f). Initially, areas with no visible corrosion can be observed, as shown in Fig. 1(d) suggesting local variations in corrosion susceptibility. These regions may result from partial passivation occurring at elevated temperatures (>600 °C). As discussed earlier, such passivation is likely incomplete; however, localized passivated zones may develop that resist liquid-metal wetting, hinder further oxygen ingress, and suppress iron dissolution. Oxygen can still diffuse through the oxide layer with longer exposure time and penetrate into the underlying matrix, leading to oxidation along GBs, as shown in Fig. 1(e). The oxidized GBs observed here are most likely pre-austenitic grain boundaries (PAGBs), based on their distinctive morphological features compared to Fig. 1(a). In some regions of the same sample, further corrosion of these GBs is observed, as shown in Fig. 1(f). The oxidation in these areas extends into the adjacent grains surrounding the corroded GBs, resulting in the formation of broader oxidized zones.

To further elucidate the underlying corrosion mechanism, compositional and crystallographic analyses were performed using EDX and EBSD, respectively. These techniques were employed to determine whether local elemental enrichment or depletion, as well as grain-orientation relationships, correlate with the observed corrosion features.

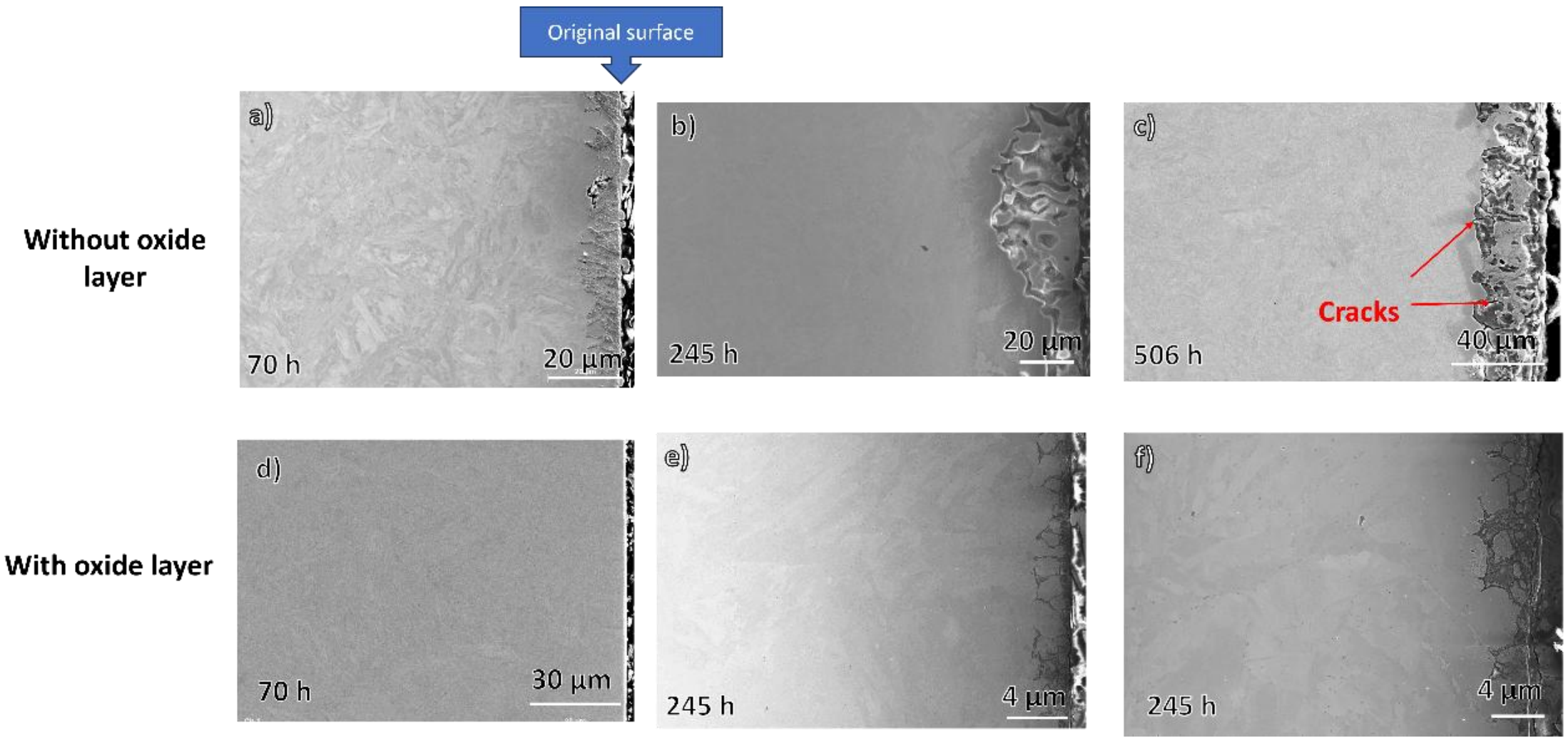

*Fig. 1. Corrosion evolution pathways of LBE-exposed samples in an oxidizing environment at 700 °C.(a–c) Corrosion without an effective protective oxide layer: (a) martensitic-lath corrosion in the sample exposed for 70 h; (b) wider-area corrosion after 245 h; (c) wider-area corrosion with visible cracks after 506 h.(d–f) Corrosion with an effective protective oxide layer: (d) region covered by a protective oxide layer showing no obvious corrosion; (e) grain-boundary corrosion observed in regions with equiaxed grains; (f) oxidation of grains adjacent to corroded grain boundaries.*

### 3.2 Intergranular internal corrosion

Fig. 1(a) and Fig. 1(e) show two different kinds of intergranular internal corrosion: in the former it follows much finer GBs, presumably associated with martensite, while in the latter corrosion appears to follow larger GBs. To study the mechanism that causes these differences, EDX and EBSD were used.

#### 3.2.1 Intergranular corrosion at martensite GBs

Fig. 2 shows SEM-EDX results of the sample corroded for 70 h. In Fig. 2 (a), intergranular internal corrosion following martensite GBs can be observed, accompanied by the formation of a surface layer over the original material. To elucidate whether regions of differing elemental concentration correlate with these features, EDX measurements were performed at 5 kV and 2 nA [23] – the low accelerating voltage allows localised measurement of EDX information with ~50 nm spatial resolution [25]. The EDX results in Figs. 2(b)–(d) captured from the blue dashed rectangular in Fig. 2(a) show that the corroded GBs contain mainly Cr oxides, with minor Si oxide formation. Although thermodynamic studies indicate that Si oxides form more readily than Fe–Cr–O oxides [21], the relatively low Si content in T91 (0.34 wt. %) compared with Cr (8.76 wt. %) may explain why Cr oxidation dominates.

The dense layer formed on the T91 surface consists primarily of Fe, with negligible Cr, Si, or O detected. As shown in Fig. 2(f), no Pb or Bi was detected in this region. To quantitatively assess the observed compositional variations, an EDX line scan was performed along the white line indicated in Fig. 2(a), with the results shown in Fig. 2(g). Three distinct regions were identified:

**Region 1**: an Fe-enriched surface layer.

**Region 2**: oxidised grain boundaries.

**Region 3:** the T91 matrix.

According to Fig. 2(g), Region 1 is predominantly Fe (>80 wt. %), with only minor amounts of other elements. To further compare the compositional features, EDX spectra were extracted, and shown in *Supplementary Materials S1*. Fe-enriched surface layer is confirmed to consist mainly of Fe. The oxygen content on the surface is too low to classify it as an Fe oxide.

Region 2 is characterised by oxides composed primarily of Cr, O, and Fe, with minor Si enrichment. The elemental ratio of O:Cr:Fe ≈ 4:2:1 is consistent with the stoichiometry of chromite-type spinel oxides reported in the literature [26, 27]. Region 3 represents the unoxidized substrate, which is depleted with Cr compared to the as-received T91 microstructure.

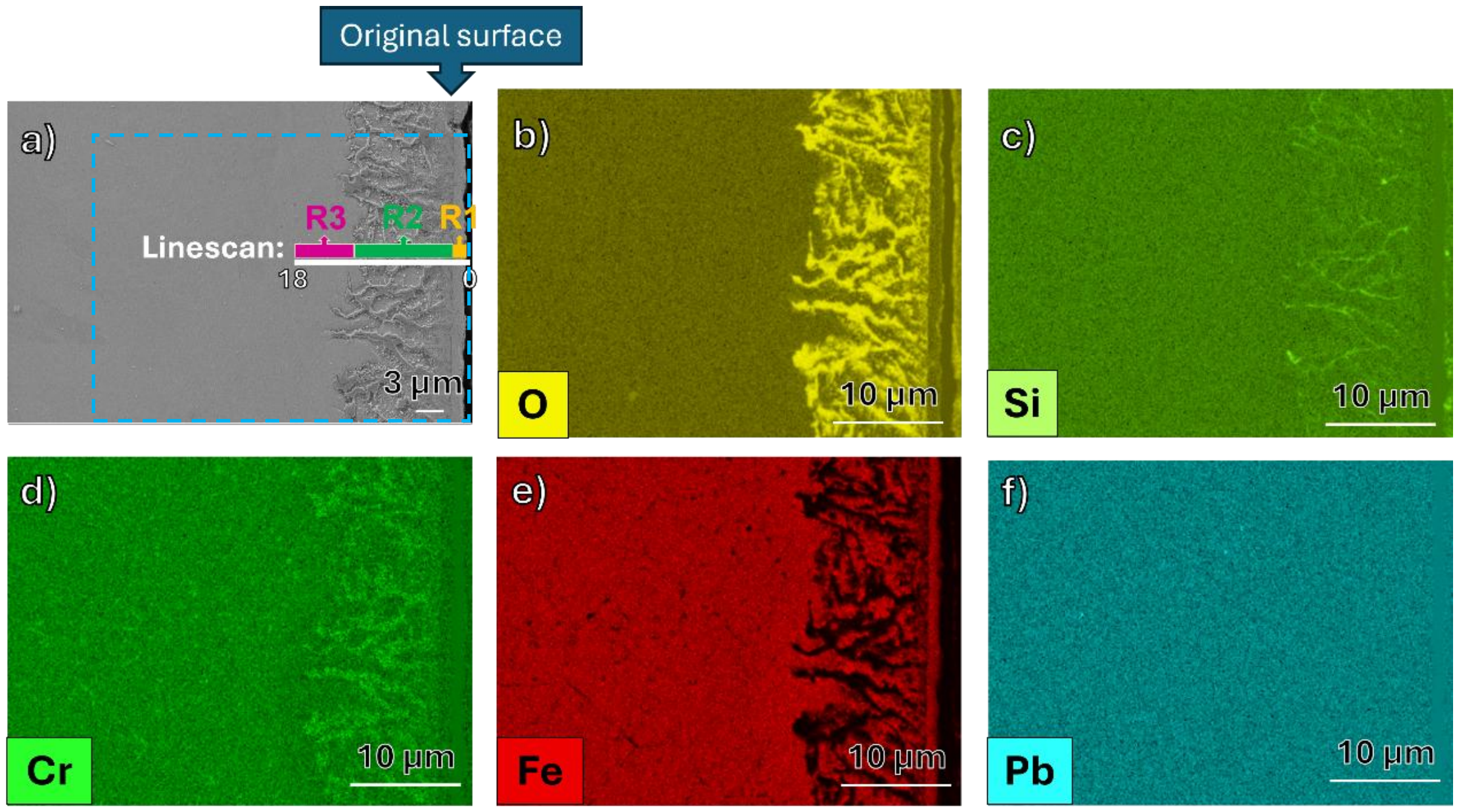

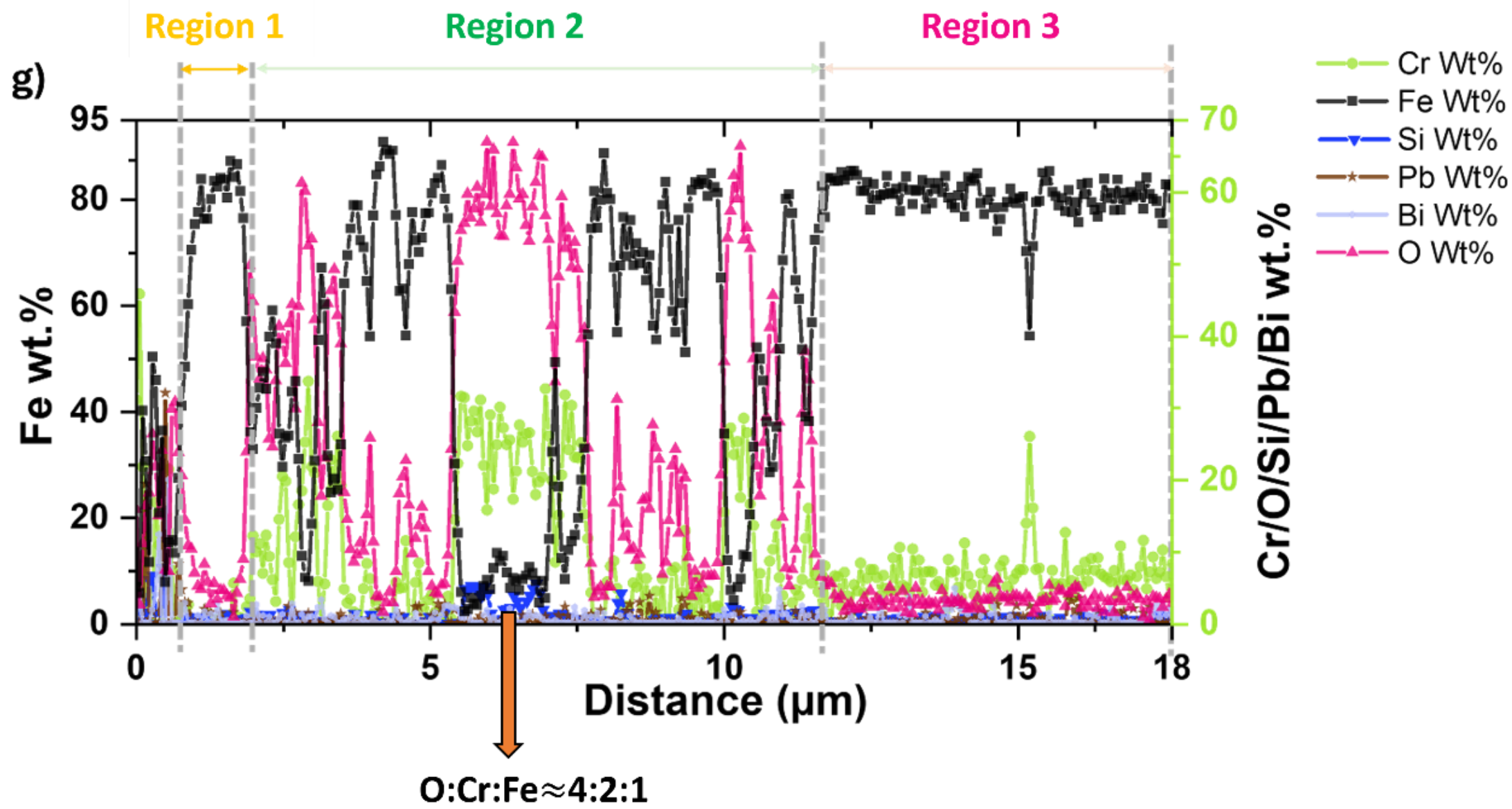

*Fig. 2. SEM-EDX [Sample: 70 h, oxidising environment, 700 ℃, LBE]. (a) SEM image with the linescan position highlighted with a white line. The different regions of the linescan are highlighted with Region 1 (R1) in orange, Region 2 (R2) in green, and Region 3 (R3) in pink. (b)-(f) EDX results for O, Si, Cr, Fe, and Pb from the blue dashed rectangle in Fig. 2(a). (g) EDX linescan along the white line shown in Fig. 2 (a) with R1, R2, and R3 marked.*

While EDX analysis provides only compositional information, determining which phase(s) exist requires knowledge of the lattice structure. Therefore, EBSD measurements were performed in a region corresponding to that shown in Fig. 2, and the results are presented in Fig. 3. The Fe-enriched surface layer is indicated by a dashed red rectangle in Fig. 3(a). As shown in Fig. 3(b), the indexed lattice structure of this layer is body-centred cubic (BCC), identical to that of the T91 matrix. Many researchers reported Fe oxides that formed outside the original sample to be chromite-type spinel oxides [26, 27]. FeO ($Fm\overline{3}m$ space group, No. 225) and $Fe_3O_4$ ($Fd\overline{3}m$ space group, No. 227) are both face-centred cubic (FCC) phases. Combining this crystallographic information with EDX results suggests that the surface layer is composed of ferrite rather than an oxide. This differs from previous reports, which commonly identified the surface layer as $Fe_3O_4$ [28, 29]. This BCC layer is spatially heterogeneous and discontinuous across the specimens examined. Additional representative data are shown in *Supplementary Materials S2 and S3* to fully document this surprising finding.

Fig. 3(c) further shows that the average grain misorientation within the oxidised GBs and the Fe-enriched surface layer is lower than in the underlying T91 matrix, suggesting microstructural relaxation during corrosion. This is well known to occur, as selective dissolution of elements allows reorganization and recrystallization of new phases which will reduce the grain misorientation. For example, Cairang et al [30-32]. recently reported ferritization of 316 stainless steel corroded in Pb-4Bi under similar conditions.

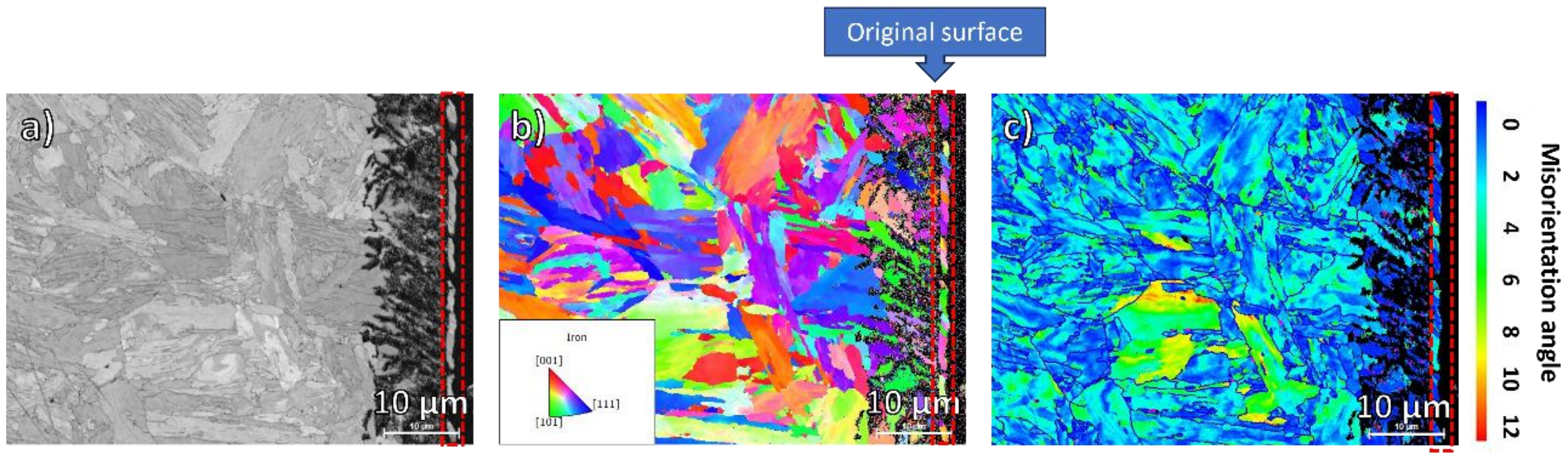

*Fig. 3. SEM-EBSD results highlighting the surface formed Fe-enriched layer with dashed rectangular [Sample: 70 h, oxidising environment, 700 ℃, LBE]. (a)* SEM image corresponding to EBSD maps*. (b) EBSD-IPFZ map, showing only material indexed as BCC. (c) EBSD grain average misorientation map.*

After a prolonged exposure of 506 h, corrosion continues to propagate preferentially along martensitic GBs. However, the extended duration allows for further elemental diffusion, leading to compositional variations in the GB oxides, as shown in Fig. 4. It is important to note that the contrast in these EDX maps is independent for each element; therefore, the brightness of a given element (e.g., Si) does not indicate its relative concentration compared to other elements. All EDX maps have been corrected for background noise but not normalized, meaning the intensity scales for different elements are not adjusted to sum to 100%.

Fig. 4(a) shows a discontinuous, Fe-enriched layer formed outside the original surface, similar to that observed in Figs. 2(a) and 3(a). Fig. 4(b) and 4(c) reveal that Cr and Si are enriched along the GBs. The EDX line scan conducted along the line indicated in Fig. 4(b) and 4(c), with results shown in Fig. 4(e), shows that Cr-rich and Si-rich oxides appear in an alternating pattern along the corroded boundary—distinct from the distribution observed in the 70 h corroded sample, which showed predominantly Cr-rich oxides. This difference likely arises because the prolonged exposure allows Si to diffuse more extensively towards the corroded GBs considering the content difference of Si and Cr in T91. No Pb and Bi were detected in the EDX map. The corresponding IPFZ map of the corroded region is presented in Fig. 4(d). Much lower accelerating voltage is needed to visualize the oxygen signal, the results can be found in *Supplementary Materials S4*.

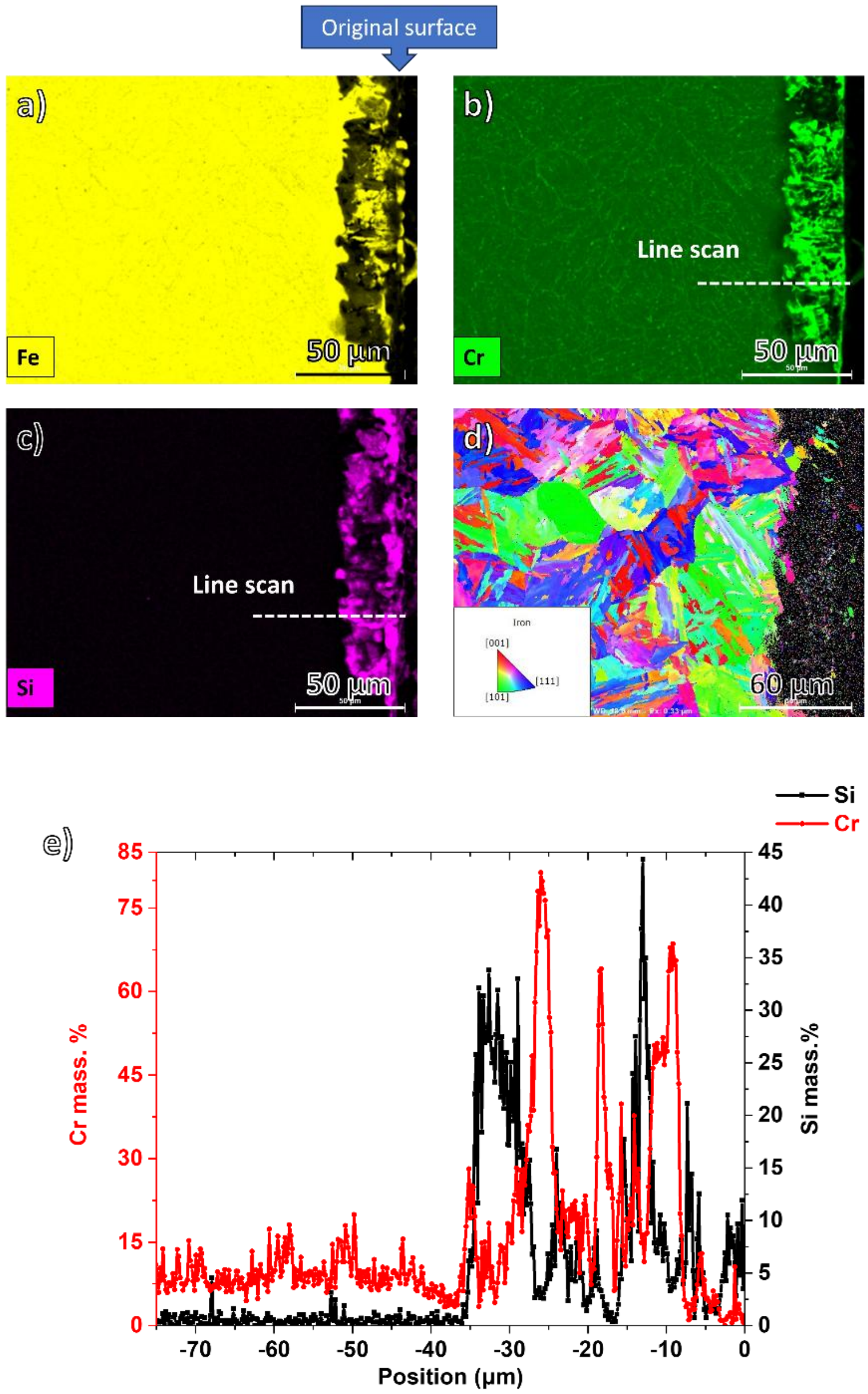

*Fig. 4. SEM-EDX and EBSD of oxidised GBs [Sample: 506 h, oxidising environment, 700 °C, LBE]. (a) EDX map highlighting Fe. (b) EDX map highlighting Cr. (c) EDX map highlighting Si. (d) EBSD-IPFZ map. (e) EDX line scan result following the line shown in Fig. 4 (b) and (c) with the original sample surface as position 0 µm.*

### 3.2.2 Intergranular internal corrosion at PAGBs

As shown in Fig. 1, the 245 h corroded sample in Fig 1(e) shows a different corrosion morphology to the 70 h sample in Fig. 1(a). In the 70 h sample, corrosion predominantly follows martensite lath boundaries. In the 245 h sample, corrosion seems to follow GBs surrounding much more equiaxed grains. To reveal the chemical composition information of the 245 h corroded sample, low energy (5 keV) EDX was used to acquire the O signal at higher spatial resolution [23]. The results are shown in Fig. 5. Combining the results shown in Fig. 5(b)-(d), it is seen that the oxidised GBs consist of Cr and Si oxides. An obvious Cr-depleted zone extending beyond the oxide layer is observed in Fig. 5(b) highlighted by a yellow dashed box. The Cr-enriched precipitates observed in the unaffected parent material on the left have dissociated in the highlighted Cr-depleted region. This is probably because Cr diffuses from the matrix to the oxidised GBs and to the surface oxide layer, leaving behind a Cr-depleted region. Such Cr depletion in the sample was also observed in the 70 h and 506 h corroded samples adjacent to the GBs oxides as shown in Fig. 2 and Fig. 4. No Pb or Bi was detected within the GB oxides, suggesting that oxidation, rather than liquid-metal penetration, governs the corrosion process here.

To quantify elemental distributions in the corroded GBs of the 245 h corroded sample, an EDX line scan was conducted through the oxidised GBs as shown in *Supplementary Materials S5*. Fe oxide, Fe-Cr oxide, and Si oxide can be observed. By analysing changes in the Cr content, a Cr depletion zone can be seen near the Cr-enriched oxides.

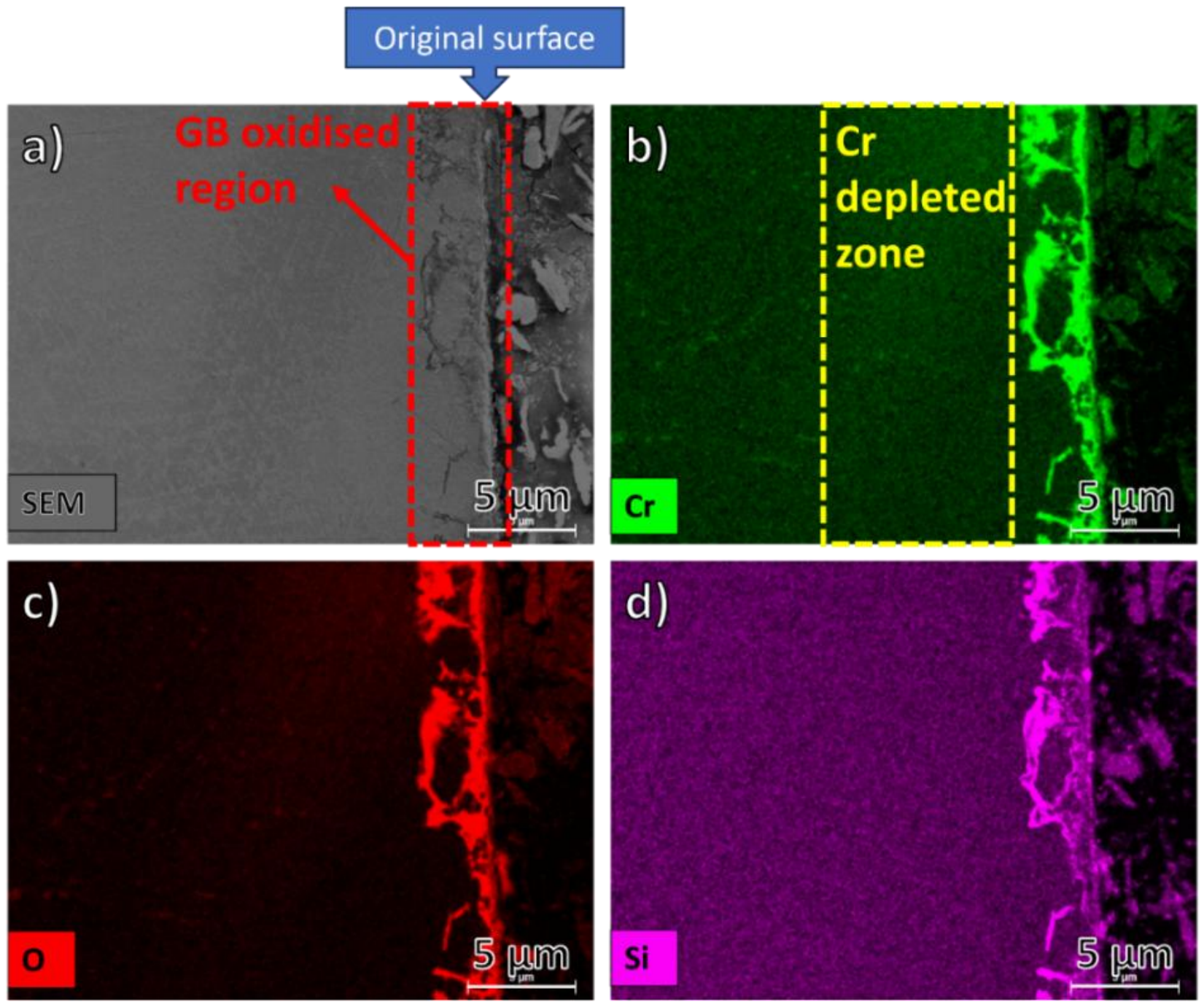

*Fig. 5. Low energy SEM-EDX results [Sample: 245 h, oxidising environment, 700 °C, LBE]. (a) SEM image with a red dashed box highlighting the GB oxidised region. (b) EDX result highlighting the Cr with the yellow dashed box showing the Cr depleted zone. (c) EDX result highlighting O. (d) EDX result highlighting Si. Pb and Bi were not detected in these maps.*

Fig. 6 shows EBSD results corresponding to the region shown in Fig. 5(a), but at a lower magnification. By comparing the scale bars of Fig. 5(a) and Fig. 6(a), the corroded surface layer, outlined by the red dashed rectangle in Fig. 6(a), can be clearly identified. The adjacent region, marked by the yellow dashed rectangle, exhibits a distinct grain morphology compared to the unaffected matrix: The areas highlighted by the red and yellow dashed rectangles in Fig. 6(b) and (c) display a transformation from martensitic laths to more equiaxed grains with lower grain-averaged misorientation. This observation suggests that martensitic decomposition has occurred in these regions.

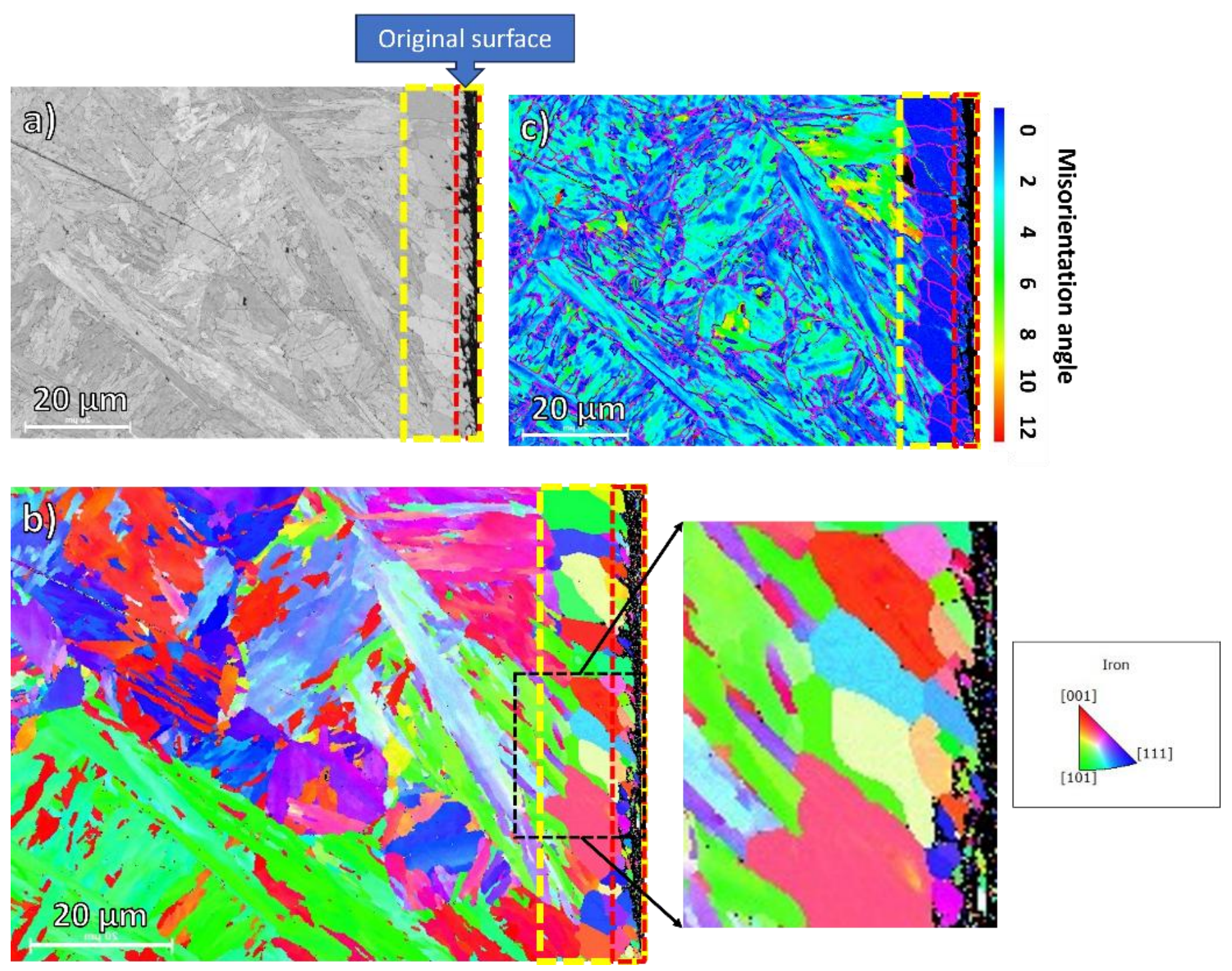

*Fig. 6. SEM-EBSD results highlighting the phase-changed region in the yellow dashed box, and the GB oxidised region in the red dashed box, corresponding to areas shown in Fig. 5 [Sample: 245 h, oxidising environment, 700 ℃, LBE]. (a) SEM result (Secondary Electron). (b) EBSD IPFZ map with the local region zoomed in for the black dashed box. (c) Grain average misorientation with scale bar.*

### 3.3 Wider area corrosion

As illustrated by Fig. 1(b) and (c), GB corrosion will eventually develop into larger area corrosion, extending beyond the original GB regions. Fig. 7 shows the initial stage of wider area corrosion following GB corrosion in the 70 h sample. EDX results in Fig. 7 were obtained at 5 kV to improve spatial resolution and provide an oxygen signal. The analysis confirms that the major corrosion products are oxides, with only trace amounts of LBE detected. Fig. 7(b)-(d) reveal the distribution of Cr- and Si-enriched oxides. In Fig. 7(e), there is an Fe depletion in the corroded region with only little Fe left. Furthermore, Fig. 7(f) shows the distribution of Mo, acquired with an accelerating voltage of 25 kV. As the Pb-M x-ray energy is 2.342 keV and the Bi-M x-ray energy is 2.419 keV, similar to the Mo-$L_\alpha$ energy of 2.293 keV, peak overlap is a concern with 5kV. *Supplementary Material S7* shows the EDX results for other elements at 25 kV. *Supplementary Material S8* also shows the EBSD results for this region.

To quantify the elemental composition and assess the extent of LBE penetration, an EDX line scan was performed along the white line indicated in Fig. 7(a), with the corresponding results presented in Fig. 7(g). Fe is depleted within the oxidized region, while Cr oxides dominate the corrosion products. Multiple Cr peaks suggest the formation of layered, chromite-type spinel oxides. Si shows only minor enrichment, and only minimal LBE penetration is observed.

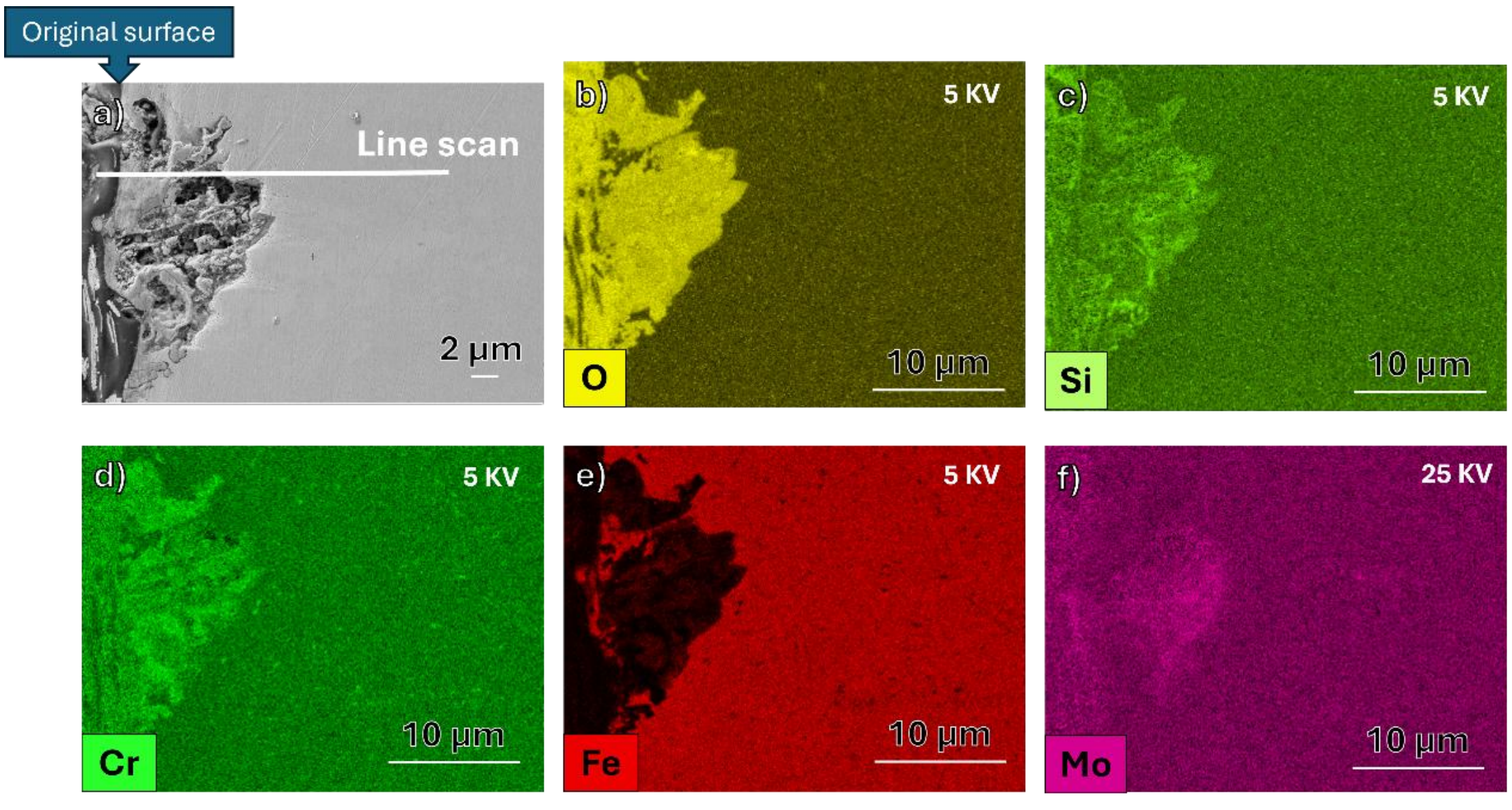

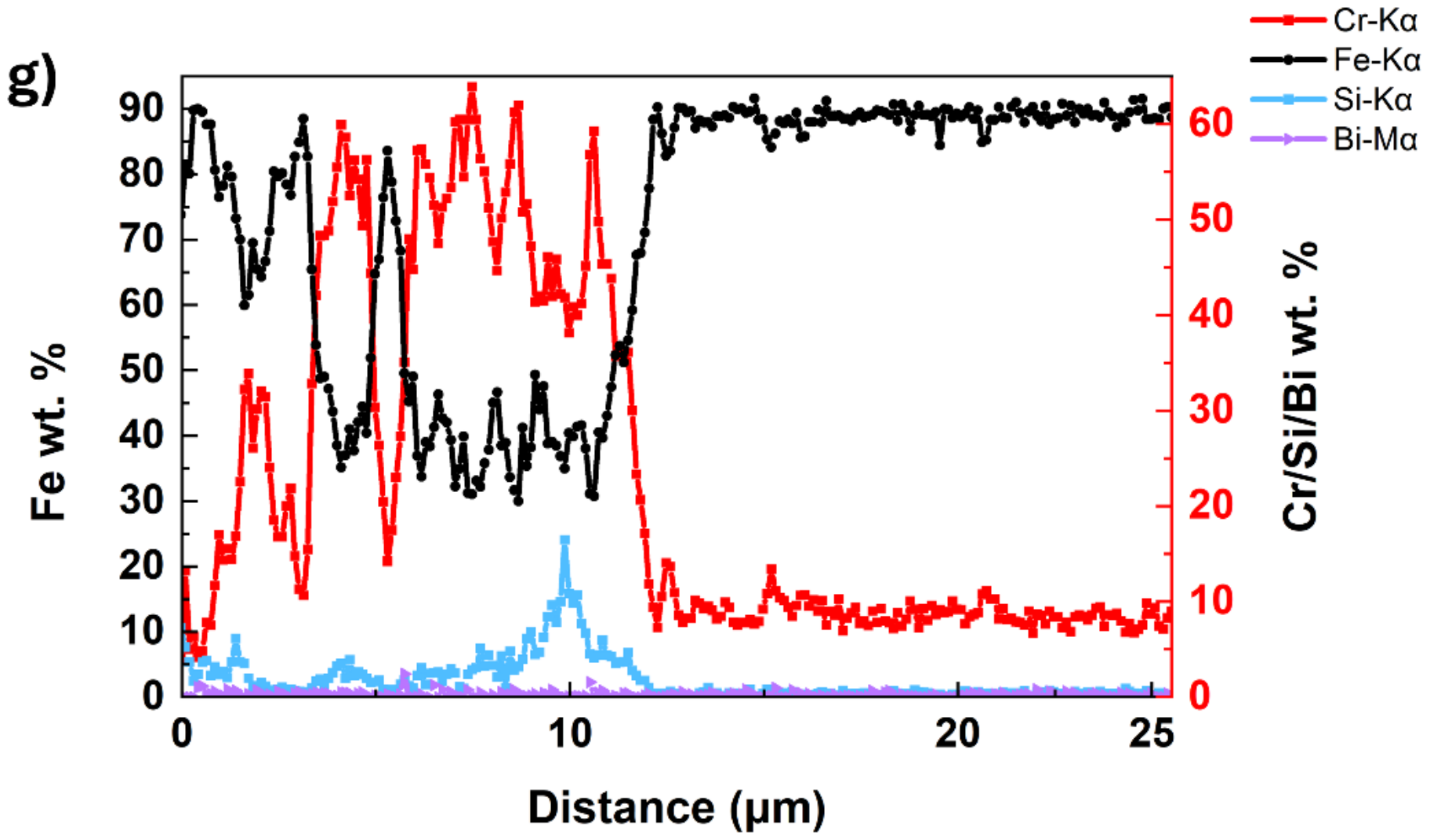

*Fig. 7. SEM-EDX results acquired at 5 kV in regions of wider area corrosion [Sample: 70 h, oxidising environment, 700 ℃, LBE]. (a) SEM image. (b)-(f) SEM-EDX results highlighting O, Si, Cr, Fe and Pb, respectively. (g) EDX line scan result for the white line in Fig. 7 (a).*

The second stage of wider area corrosion occurs after oxide formation, when the oxide layer detaches from the surface, allowing LBE to further penetrate the material. Fig. 8 shows one such region exhibiting more severe LBE penetration in the 245 h sample. Fig. 8(d) and (e) reveal that corrosion at this stage is dominated by LBE intrusion, characterised by high concentrations of Pb and Bi. This may account for the low Cr content observed in Fig. 8(b), as LBE likely dissolves the surrounding Cr during penetration due to its relatively high solubility compared to other alloying elements in T91. There is a discontinuous Cr-enriched oxide layer at the surface, which appears insufficient to hinder further LBE penetration.

Fig. 8(c) shows Si enrichment around the LBE intrusions, where Si-enriched oxides form adjacent to LBE-rich regions, with almost no chromite-type spinel oxides nearby. This suggests that Si-enriched oxides are more resistant to LBE corrosion than other oxides. Additionally, Fig. 8(f) shows a discontinues Fe-enriched surface layer, likely formed through outward Fe diffusion as mentioned in literature [28].

To further verify the elemental distribution, a line scan was performed along the white line indicated in Fig. 8(a) and the results are shown in *Supplementary Material S9*.

The EDX result with 5 KV accelerating voltage for wider area corrosion region is shown in *Supplementary Materials S10* to reveal the O signal. Lower accelerating voltage helps to reveal the O signal but at the cost of a lower signal to noise ratio than in higher energy scans.

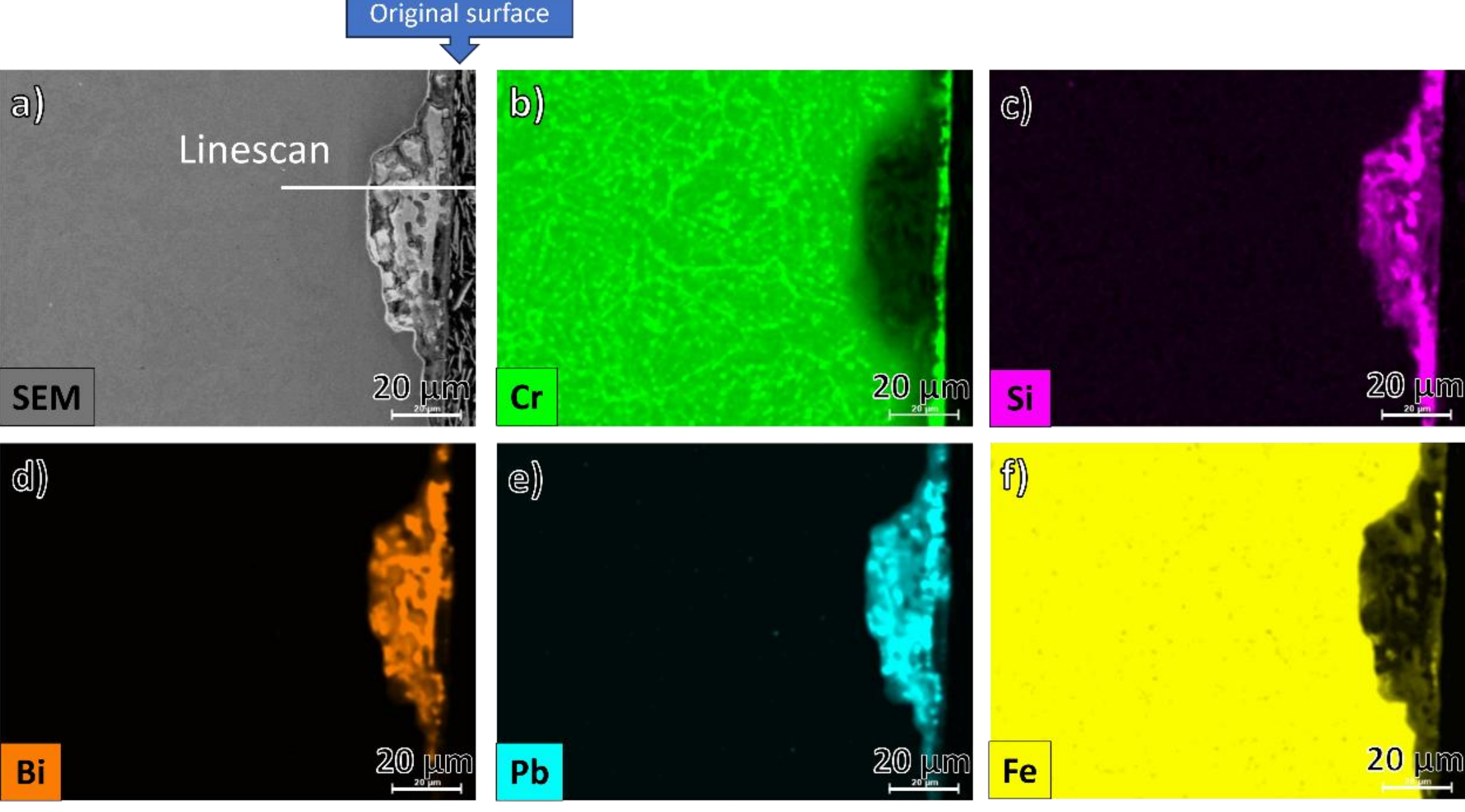

*Fig. 8. SEM-EDX and EBSD results acquired at 20 kV [Sample: 245 h, oxidising environment, 700 ℃, LBE]. (a) SEM view indicating the position of the EDX line scan in Fig. 8(g) highlighted by a white line. (b) (c) (d) (e)(f) SEM-EDX results highlighting Cr, Si, Bi, Pb, and Fe, respectively.*

### 3.4 Cracks between oxide and substrate

Cracks, potentially caused by the mismatch in thermal expansion between the substrate and the oxide scale, are observed in samples corroded for 245 h and 506 h. Crack formation becomes increasingly probable with increasing oxide thickness due to the progressive accumulation of strain energy within the growing oxide scale. Because longer exposure times promote thicker oxide layers, they lead to higher stored strain energy and, consequently, a greater propensity for cracking [18]. Accordingly, a higher crack density is observed at longer exposure durations. As shown in Fig. 9, intergranular corrosion along GBs can be observed in the regions highlighted by the black dashed boxes. Fig. 9(f) further reveals that Pb penetrates through these cracks, enabling its penetration into the deeper layers of the material and thereby exacerbating corrosion. Corrosion-induced cracking can significantly reduce the mechanical strength of the material and may lead to oxide spallation within the circulation system, potentially resulting in blockage. Similar cracking behaviour observed in the 506 h corroded sample is shown in *Supplementary Material S11.*

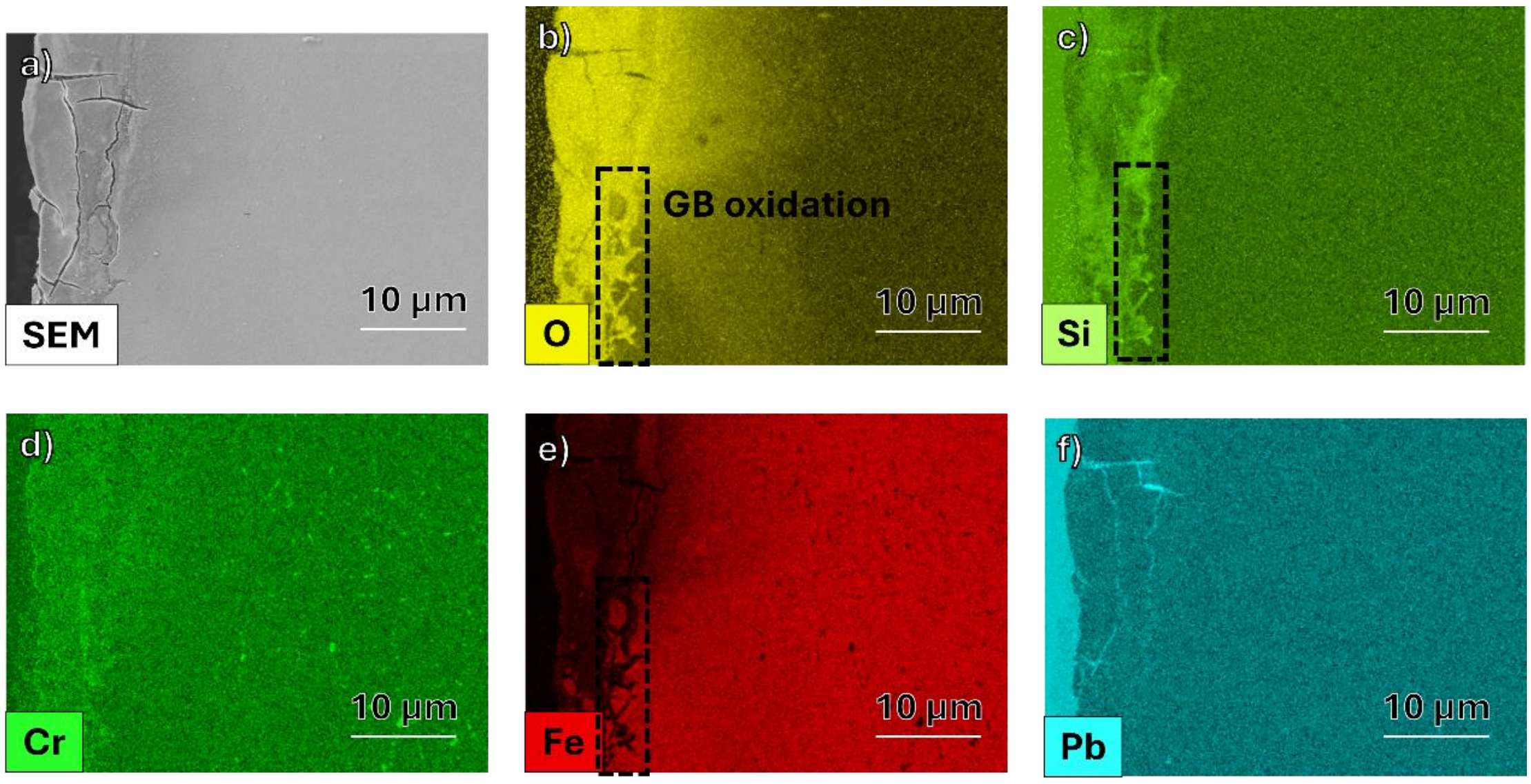

*Fig. 9. SEM-EDX with cracks observed in the oxide layer [Sample: 245 h, oxidising environment, 700 ℃, LBE]. (a) SEM image. (b)-(f) EDX results highlight O, Si, Fe, Cr, and Pb, respectively, with GB oxidation highlighted.*

### 3.5 Regions covered by an effective protective surface layer

All samples of different corrosion duration time contain regions showing no apparent signs of corrosion, as illustrated in Fig. 1(d). Studying these unaffected areas is crucial, as they may provide insights into how the corrosion resistance of T91 steel may be enhanced.

Fig. 10 shows the SEM-EDS and EBSD result for the 506 h corroded sample. There is no visible oxidation or corrosion in Fig. 10(a), and a lack of Fe depletion in Fig. 10(d) further confirms this. Fig. 10(b) and (c) confirm the formation of a dense layer of Si and Cr enriched oxide. It can be inferred that the formation of a Cr- and Si-enriched surface oxide layer effectively protected the underlying material. Fig 10(b) shows a Cr depleted region adjacent to the oxide layer which may have formed because Cr diffused to the surface to form the oxide layer. Fig. 10(d) shows that there is negligible Fe in the protective oxide layer. To further prove the chemical composition of the protection layer, the SEM-EDX result of 70 h corroded sample is shown in *Supplementary Materials S12*.

The corresponding EBSD results in Fig. 10(e) and (f) show more equiaxed grains in this region, accompanied by a significantly lower grain-averaged misorientation, similar to that observed in Fig. 5 and Fig. 6. These observations suggest that Cr depletion may have induced martensitic decomposition in this region. The amount of Cr depletion varies spatially and correlates with the level of martensitic decomposition. This is confirmed by *Supplementary Material S13*, which shows another region that exhibits no identifiable corrosion, similar to Fig. 10. Here, less Cr depletion, and accordingly a smaller phase changed layer is observed.

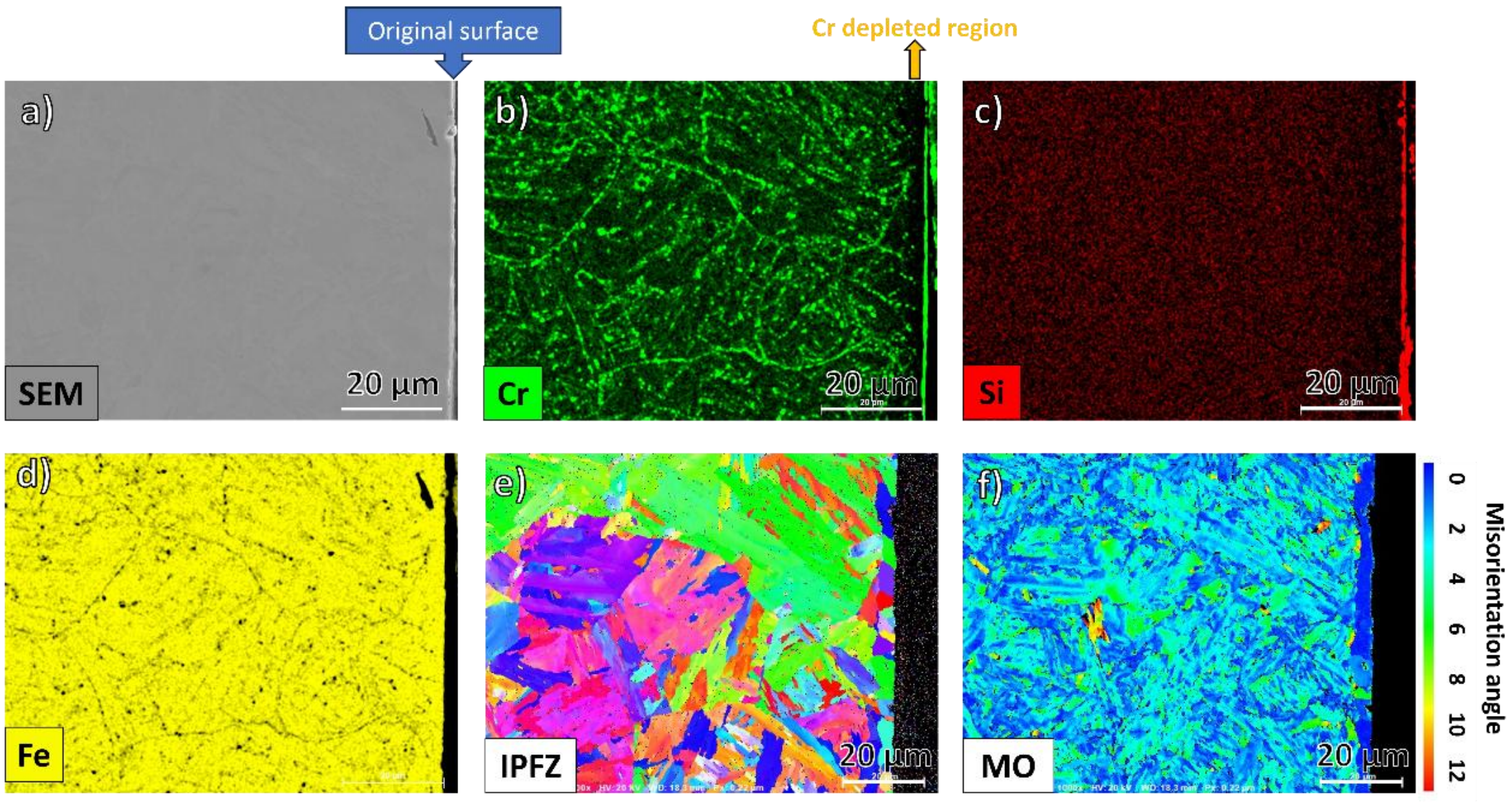

*Fig. 10. SEM-EDX and EBSD results for one representative area with no obvious corrosion [Sample: 506 h, oxidising environment, 700 ℃, LBE]. (a) SEM SE view. (b) SEM-EDX result highlighting Cr with the Cr depleted region marked. (c), (d) SEM-EDX*

*result highlighting Si and Fe, respectively. (e) SEM-EBSD IPFZ map. (f) SEM-EBSD grain average misorientation (MO) with scale bar.*

## 4. Discussion

In this study, T91 samples exposed to LBE under oxidizing conditions for 70 h, 245 h and 506 h at 700 ℃ were analyzed. The oxides located at different positions of the oxide film have been examined in using EDX and EBSD to understand the evolution of both chemical composition and crystallography.

Two distinct corrosion evolution pathways were identified depending on whether an effective protective oxide layer formed on the surface. In regions where such a protective layer developed, corrosion began with the formation of a dense surface oxide, followed by oxidation along PAGBs. In contrast, regions lacking an effective protective oxide layer exhibited initial corrosion along martensitic lath or GBs, which subsequently progressed into wider-area corrosion.

In oxidized GBs we observed 2 forms of products: A Fe/steel BCC layer outside the original sample surface and a Cr/Si enriched discontinuous oxide layer. The wider area corrosion showed extensive oxide formation in some regions and LBE penetration after prolonged exposure. In regions without obvious corrosion, a dense Cr/Si oxide layer is observed. Based on the results presented above and in literature, detailed formation mechanisms of the different oxide layers are developed below.

### 4.1 Formation of GB oxides

There is a substantial body of literature on the formation of oxides layers on F/M steels due to exposure to high temperature media, such as steam [26, 27], carbon dioxide [33], and liquid Pb or LBE [34, 35]. All these studies propose formation and growth mechanisms for an outer oxide layer which grows outside the original surface, and an inner layer (inside the original steel surface). The oxidation behaviour of T91 steel is governed by elemental diffusion in both T91 and LBE. At the T91 - liquid LBE interface, there are 2 possible diffusion paths: (a) elements from T91 diffuse outwards to liquid LBE and form oxides; (2) oxygen diffuses inwards from the liquid LBE into T91 to form oxides.

The oxygen content fluctuations and the oxide morphology in Fig. 2(g) suggest the diffusion of oxygen into T91, i.e. an inward diffusion process. This is consistent with literature observations [34, 35]. The significantly more negative Gibbs free energy associated with formation of $Cr_2O_3$ and $SiO_2$ compared to Fe-oxide phases drives the preferential oxidation of Cr and Si [36-38]. According to the results reported in the literature[26, 39, 40] , with the increase of environmental oxygen partial pressure, the formation of different types of oxides in Fe-Cr steels follows the sequence of $Cr_2O_3$, $FeCr_2O_4$, $Fe_3O_4$, and $Fe_2O_3$. As a result, oxygen diffusing inward reacts primarily with these solutes, leading to the formation of Cr and Si enriched oxides and Fe-Cr-O spinel-type oxides within the alloy.

Interestingly all of the above studies [26, 27, 33-35] claim that oxygen diffusion is the main factor controlling the formation of Cr-enriched oxides, as the Cr concentration in the oxide remains similar to that observed in the unaffected matrix, i.e. there is little diffusion/clustering of Cr. This is very different to our observations in Fig. 2(g), Fig. 4(e), and Fig. 7(g). A much higher Cr content is observed in oxidised GBs than in the unaffected matrix. Furthermore, our observations show the depletion of Cr in the vicinity of the oxides. The key difference is the test temperatures: the studies above were conducted below 600 ℃, while our experiments were carried out at 700 ℃. A higher temperature may allow Cr to diffuse from surrounding material to the corroded GBs. Ye's research at 823 K also showed a Cr enriched zone in oxides and a Cr depleted zone in the vicinity [41]. Chen's study at 790 ℃ similarly reported a Cr enrichment in the oxide area [27]. Thus, we conclude that at temperatures of 700 ℃ Cr appears to be sufficiently mobile that it can diffuse during the oxidation process. More generally, our observations suggest that at higher temperatures both the diffusion of alloying elements and oxygen control the corrosion process.

**4.2 Fe-rich layer on surface**

In several previous studies the structure of the outer oxide layer and its formation mechanism have been discussed [27, 34, 35]. It is believed that the formation of the outer oxide layer is due to the outward diffusion of Fe from the inner oxide layers, leading to the formatting of a Fe-dominated outer oxide layer [26, 27].Different types of outer oxides have been reported in the literatures, the most common ones are $Fe_2O_3$ and $Fe_3O_4$. The oxygen partial pressure predominantly determines which oxide is formed: High oxygen partial pressure favours the formation of $Fe_2O_3$ while a lower pressure will lead to the formation of $Fe_3O_4$ [42, 43].

Remarkably, in our study, the outer layer observed in the 70 h corroded sample mainly consists of Fe (over 80 wt. %) and exhibits a BCC lattice structure, as shown in Figs. 2 and 3. Since $Fe_2O_3$ possess hexagonal structure, and $Fe_3O_4$ /FeO both have FCC structure, and no new phases are observed in the EBSD phase map, the outer layer shown in Figs. 2 and 3 cannot be attributed to either $Fe_2O_3$ or $Fe_3O_4$. Instead, it is most likely to be a ferritic steel phase with a high Fe content. There are 2 possible explanations. A potential explanation could be that initially $Fe_3O_4$ is formed, which subsequently loses its oxygen to Cr and Si as oxygen diffuses into the sample. This would require the uptake of oxygen from the environment (LBE) to be significantly lower than the rate of oxygen removal due to the formation of Cr and Si oxides. Consistent with this hypothesis, the outermost Fe layer in Fig. 2(b) still shows some O enrichment, so a very thin layer of $Fe_2O_3$ or $Fe_3O_4$ may still exist. According to the literature, $Fe_2O_3$ and $Fe_3O_4$ outer oxide layer is generally believed to be non-protective with cavities and micro-cracks [44, 45]. Another possible mechanism is that Fe, dissolved in the LBE, becomes redeposited onto the surface, forming an Fe-enriched layer. A thin oxide layer subsequently develops on top of this deposition layer due to oxidation occurring after the redeposition process.

### 4.3 Mechanisms and evolution of the corrosion process

Different types of intergranular internal corrosion can occur depending on the morphology of the surface oxide layer. In areas without a protective surface oxide, martensite GBs will act as fast oxygen diffusion paths, leading to GB oxidation as shown in Fig. 1(a). In areas with a protective surface oxide, oxygen will diffuse into the material more slowly [46]. The formation of an oxide layer on the material surface will consume Cr, drawing Cr out of precipitates and the material matrix, as seen in Fig. 5(b), and Fig. 10(b). The lower Cr content leads to a reduced recrystallisation temperature [47], meaning that grains with lower Cr content start to recrystallise at the test temperature (700 ℃) while the grains with higher Cr content remain stable. As such this removal of Cr will then trigger the recrystallization of martensite to ferrite as seen in Fig. 6 and Fig. 10(e) and (f). During this reversion to ferrite, the grains will change from the martensitic lath structure to more equal-axis grains [13]. After oxygen penetrates through the oxide layer, it travels along the newly formed ferrite GBs as fast diffusion paths and drive GB oxidation as shown in Fig. 1(e).

The corrosion morphology shown in Fig. 1(a)-(c) suggests a progression from predominantly intergranular internal oxidation to wide area corrosion. As shown in Fig. 7 and Fig. 8, the degradation evolves from the formation of coarse internal oxides to pronounced liquid-metal (LBE) penetration within the grains. Ye [41] proposed the following mechanism: The outwards diffusion of Fe gives rise to high concentration of vacancies in the steel matrix which accumulate to form cavities. At 700 °C, vacancies in steel are expected to be highly mobile [48]. The cavities connect to surfaces, allowing oxygen to diffuse in and form oxides at the edges (primarily Cr oxide). This will lead to the formation of Cr depleted regions and stop further oxidation until more oxygen accumulates in the cavities and Cr from the unaffected matrix has diffused to the depleted region. Increase of the chemical potential of O and Cr gradually makes oxidation possible in Cr depleted regions. Progression of this process will result in the wide area oxidation as shown in Fig. 7.

As the internal oxides thicken, the mismatch in volume between the oxide layer and the steel substrate generates local stresses [49]. The volume change due to oxidation promotes cracking and eventual spallation of the oxide layer, as evidenced in Fig. 9 and *Supplementary Materials S11*. Once the oxide fractures or detaches, fresh steel is exposed to LBE, allowing further penetration along GBs and eventually into grain interiors. This repeated cycle of internal oxidation, dissolution by LBE, and oxide cracking produces the characteristic subsurface morphology observed in Fig. 8.

Through this repeating process, LBE progressively consumes alloying elements and disrupts the microstructure, reducing the load-bearing capacity and long-term integrity of T91. The observed transition from intergranular corrosion to wider area corrosion thus reflects the combined effects of rapid oxygen ingress, internal oxidation, liquid metal dissolution, and mechanically induced oxide degradation.

**4.4 Phase transformation**

Phase transformations are observed in both Fig. 6 and Fig. 10(d-e), and in each case, they coincide with pronounced local Cr depletion, as shown in Fig. 5(b) and Fig. 10(b). This relationship is further supported by comparing Fig. 10 with *Supplementary Material S13*, both obtained from the same sample. The region shown in *S13* exhibits a substantially smaller Cr-depleted zone, where only subtle phase changes are visible. This comparison indicates that a larger Cr-depleted region corresponds to a more extensive microstructural transformation.

Previous studies [13] have demonstrated that Cr depletion reduces the recrystallization temperature of the near-surface region in T91 F/M steel. At the corrosion temperature used in this work (700 °C), the Cr-depleted martensitic laths therefore undergo recovery and recrystallization to form equiaxed ferrite grains, which subsequently coarsen through grain growth. During recovery, residual dislocation structures are annihilated, internal stresses decrease, and lattice misorientation is significantly reduced. The carbon released from the dissolution of Cr-rich carbides (as shown in Fig. 5(b) and Fig. 10(b)) does not hinder this ferritization process because its concentration remains below the solubility limit of C in ferrite at 700 °C.

**4.5 Uncorroded regions**

The uncorroded regions consistently exhibit a continuous and adherent oxide layer on the surface. The presence of a dense, protective oxide scale—typically Cr-rich—is known to suppress both oxygen ingress and liquid-metal dissolution, thereby preventing microstructural degradation beneath the surface [37, 38]. In addition, the wettability of oxides by LBE is far lower than that of bare metals, stopping the initiation of any dissolutive liquid metal corrosion mechanisms in these fully passivated regions. Maintaining such a continuous oxide layer is therefore critical for ensuring long-term corrosion resistance of T91 in LBE environments. Achieving this requires sufficient Cr availability at the surface to sustain a Cr-enriched oxide formation, controlled oxygen potential in the LBE to avoid under- or over-oxidation, and microstructural conditions that minimize oxide cracking or spallation [50, 51]. Recent studies further emphasize that alloy design (eg: optimized Cr content, Si and Al added), surface treatments (coatings), and precise oxygen control are essential strategies for stabilizing protective oxide scales under high-temperature LBE exposure [52, 53]. Thus, the ability to retain a continuous, mechanically stable oxide layer emerges as a central criterion for future improvements in corrosion resistance.

## 5. Conclusion

Based on the mechanistic analysis presented in the discussion, the oxidation and corrosion behaviour of T91 exposed to oxygen-controlled LBE at 700 °C can be described by the following conclusions:

The evolution process is determined by whether a protective layer forms on the surface.

**Corrosion follows GBs:**

The corrosion process begins with intergranular internal corrosion, with Cr-enriched and Si-enriched oxides forming along grain boundaries. In the reducing environment, the dissolution of elements (Cr, Ni, Fe…) plays the main role in the corrosion, which induces a corrosion pattern that does not follow the GBs. In an oxidising environment, oxygen plays an important role. The corrosion process begins with oxidation, and this process is controlled by both Cr diffusion and oxygen diffusion in T91.

**Fe displacement and Fe-enriched layer formation:**

A Fe enriched layer with BCC lattice structure was observed outside the original surface. EDX and EBSD results suggest that this layer is ferritic and consist mainly of Fe. We hypothesise that this layer is formed when an initial Fe-oxide layer loses oxygen to internal Cr/Si oxides or redeposition of Fe in LBE to the surface.

**Corrosion transitions from intergranular internal attack to wider area corrosion**

Outward diffusion of Fe leaves behind a high concentration of vacancies which accumulate into nano cavities. These cavities act as fast permeation paths for oxygen, which allows further oxidation of internal areas. With Cr diffusing from the matrix and oxygen potential increasing, the localized oxidation will develop into wider area corrosion. Volumetric mismatch between the thickening oxide and the steel substrate will generate mechanical stresses that ultimately cause oxide cracking and spallation. Once the oxide layer fractures, LBE can penetrate into grain interiors. Here it dissolves Fe, Cr, Si in rough proportion to their solubilities in LBE, and accelerates subsurface degradation. The interplay of internal oxidation, dissolution, and repeated oxide fracture results in progressively wider area corrosion.

**Cr depletion induces microstructural transformation from martensite to ferrite.**

Regions with severe Cr depletion experience significantly reduced recrystallisation temperatures. At 700 °C, the Cr-depleted martensite undergoes recovery, decomposition, and recrystallisation into equiaxed ferrite grains, accompanied by reduced lattice misorientation and defect density. Carbon released from the dissolution of Cr-rich carbides remains below its solubility limit in ferrite and does not inhibit ferritization.

**Continuous protective oxide layers are essential for corrosion resistance.**

Unaffected regions consistently show stable, adherent Cr-rich oxide scales that suppress both oxygen ingress and LBE-driven dissolution. Maintaining such protective scales requires adequate Cr availability, controlled oxygen potential in LBE, and microstructures that resist oxide cracking and spallation. These findings highlight oxide-scale stability as a critical design parameter for improving long-term T91 performance in LBE systems.

## Statement on the Use of Generative AI

During the preparation of this work, the authors used ChatGPT for English language correction and improvement of expression. After using this tool, the authors carefully reviewed and edited the content as needed and take full responsibility for the accuracy, originality, and integrity of the publication.

## Credit Authorship Contribution Statement

**Minyi Zhang:** Writing – original draft, Methodology, Investigation, Formal analysis, Conceptualization.

**Zhou Weiyue:** Writing – review & editing, Methodology.

**Michael P. Short:** Writing – review & editing, Supervision, Methodology, Conceptualization.

**Paul A.J. Bagot:** Supervision.

**Michael P. Moody:** Supervision, Methodology, Conceptualization.

**Felix Hofmann:** Writing – review & editing, Supervision, Methodology, Conceptualization.

## Declaration of competing interest

The authors declare that they have no known competing financial interests or personal relationships that could have influenced the work reported in this paper.

## Acknowledgements

EP/T002808/1, Simultaneous Corrosion/Irradiation Testing in Lead and Lead-Bismuth Eutectic: The Radiation Decelerated Corrosion Hypothesis (RC-3).

EP/T011505/1, An Atomic-Scale Characterisation Facility for Active Nuclear Materials.

NEUP 19-16754, Simultaneous Corrosion/Irradiation Testing in Lead and Lead-Bismuth Eutectic: The Radiation Decelerated Corrosion Hypothesis.

EP/R010145/1, The authors acknowledge use of characterisation facilities within the David Cockayne Centre for Electron Microscopy, Department of Materials, University of Oxford, alongside financial support provided by the Henry Royce Institute.

## Data availability

The original data from this study are freely available at:

https://ealga.mit.edu/Permanent/2026_T91_Corrosion_Oxidizing_Environment/

## References:

1. Gong, X., et al., *Atomic-scale dissolution corrosion mechanism of additively-manufactured 316L steels in liquid lead-bismuth eutectic*. Acta Materialia, 2025. **290**: p. 120963.
2. Zhang, M., et al., *Nano-scale corrosion mechanism of T91 steel in static lead-bismuth eutectic: A combined APT, EBSD, and STEM investigation*. Acta Materialia, 2024: p. 119883.
3. Davis, T.P., *Dispelling misconceptions of nuclear energy technology: How Generation IV nuclear reactors could become the key to achieving the Paris Agreement and the United Kingdom's net zero CO2 emissions target by 2050*. Saint Anne's Academic Review, 2019. **9**.
4. Garner, F.A., M.B. Toloczko, and B.H. Sencer, *Comparison of swelling and irradiation creep behavior of fcc-austenitic and bcc-ferritic/martensitic alloys at high neutron exposure*. Journal of Nuclear Materials, 2000. **276**(1-3): p. 123-142.
5. Davis, T.P., et al., *Atom probe characterisation of segregation driven Cu and Mn–Ni–Si co-precipitation in neutron irradiated T91 tempered-martensitic steel*. Materialia, 2020. **14**.
6. Kohyama, A., et al., *Low-activation ferritic and martensitic steels for fusion application*. Journal of Nuclear Materials, 1996. **233**: p. 138-147.
7. Kurata, Y. and S. Saito, *Temperature Dependence of Corrosion of Ferritic/Martensitic and Austenitic Steels in Liquid Lead-Bismuth Eutectic*. Materials Transactions, 2009. **50**(10): p. 2410-2417.
8. Alemberti, A., et al., *Overview of lead-cooled fast reactor activities*. Progress in Nuclear Energy, 2014. **77**: p. 300-307.
9. Rebak, R.B. and D.D. Ellis. *Passivation Characteristics of Ferritic Stainless Materials in Simulated Reactor Environments*. in *NACE CORROSION*. 2016. NACE.
10. Li, N., *Active control of oxygen in molten lead–bismuth eutectic systems to prevent steel corrosion and coolant contamination*. Journal of Nuclear materials, 2002. **300**(1): p. 73-81.
11. Popovic, M.P., et al., *Oxidative passivation of Fe–Cr–Al steels in lead-bismuth eutectic under oxygen-controlled static conditions at 700° and 800° C*. Journal of Nuclear Materials, 2019. **523**: p. 172-181.
12. Weisenburger, A., et al. *Stability of oxide layer formed on high-chromium steels in LBE under oxygen content and temperature fluctuation*. in *The 13th international conference on nuclear engineering abstracts*. 2005.
13. Zhang, M., et al., *Correlated chromium carbide dissociation and phase transformation in liquid lead-bismuth eutectic corroded T91 steel*. Corrosion Science, 2025. **249**: p. 112851.
14. Courouau, J.-L., et al., *Impurities and oxygen control in lead alloys*. Journal of Nuclear Materials, 2002. **301**(1): p. 53-59.
15. Barbier, F., et al., *Compatibility tests of steels in flowing liquid lead–bismuth*. Journal of Nuclear Materials, 2001. **295**(2-3): p. 149-156.
16. Tas, H., et al., *Liquid breeder materials*. Journal of Nuclear Materials, 1988. **155**: p. 178-187.
17. Simon, N., A. Terlain, and T. Flament, *The compatibility of austenitic materials with liquid Pb–17Li*. Corrosion science, 2001. **43**(6): p. 1041-1052.

18. Laverde, D., T. Gomez-Acebo, and F. Castro, *Continuous and cyclic oxidation of T91 ferritic steel under steam*. Corrosion science, 2004. **46**(3): p. 613-631.
19. Was, G., et al., *Corrosion and stress corrosion cracking in supercritical water*. Journal of Nuclear Materials, 2007. **371**(1-3): p. 176-201.
20. Guntz, G., et al., *The T91 book*. Vallourec Industries, 1990.
21. Short, M., R. Ballinger, and H. Hänninen, *Corrosion resistance of alloys F91 and Fe–12Cr–2Si in lead–bismuth eutectic up to 715 C*. Journal of nuclear materials, 2013. **434**(1-3): p. 259-281.
22. Short, M.P., *The Design of a functionally graded composite for service in high temperature lead and lead-bismuth cooled nuclear reactors*. 2010, Massachusetts Institute of Technology.
23. Meisnar, M., et al., *Low-energy EDX–A novel approach to study stress corrosion cracking in SUS304 stainless steel via scanning electron microscopy*. Micron, 2014. **66**: p. 16-22.
24. Shreir, L.L., *Corrosion: metal/environment reactions*. 2013: Newnes.
25. Berger, M.J., Coursey, J.S., Zucker, M.A., and Chang, J. *ESTAR, PSTAR, and ASTAR: Computer Programs for Calculating Stopping-Power and Range Tables for Electrons, Protons, and Helium Ions*. NIST Standard Reference Database 124 2017 2.2.2026]; Available from: https://physics.nist.gov/PhysRefData/Star/Text/ESTAR.html?utm_source=chatgpt.com.
26. Shen, Z., et al., *New insights into the oxidation mechanisms of a Ferritic-Martensitic steel in high-temperature steam*. Acta Materialia, 2020. **194**: p. 522-539.
27. Chen, K., L. Zhang, and Z. Shen, *Understanding the surface oxide evolution of T91 ferritic-martensitic steel in supercritical water through advanced characterization*. Acta Materialia, 2020. **194**: p. 156-167.
28. Shen, Z., et al., *New insights into the oxidation mechanisms of a Ferritic-Martensitic steel in high-temperature steam*. Acta Materialia, 2020. **194**: p. 522-539.
29. Chen, G., et al., *Ultrastable lubricating properties of robust self-repairing tribofilms enabled by in situ-assembled polydopamine nanoparticles*. Langmuir, 2020. **36**(4): p. 852-861.
30. Cairang, W., et al., *Simultaneous proton irradiation and dissolution corrosion of SS316L in liquid Pb-4Bi alloy*. Corrosion Science, 2025: p. 113010.
31. Zhou, Q., Z. Zheng, and Y. Gao, *Abnormal selective dissolution by the partial recrystallization in a plastically deformed austenitic stainless steel*. Corrosion Science, 2021. **188**: p. 109548.
32. de Souza Silva, E.M.F., G.S. da Fonseca, and E.A. Ferreira, *Microstructural and selective dissolution analysis of 316L austenitic stainless steel*. Journal of Materials Research and Technology, 2021. **15**: p. 4317-4329.
33. Martinelli, L., et al., *Comparative oxidation behaviour of Fe-9Cr steel in CO2 and H2O at 550 C: Detailed analysis of the inner oxide layer*. Corrosion Science, 2015. **100**: p. 253-266.
34. Martinelli, L., et al., *Oxidation mechanism of a Fe–9Cr–1Mo steel by liquid Pb–Bi eutectic alloy (Part I)*. Corrosion Science, 2008. **50**(9): p. 2523-2536.
35. Martinelli, L., et al., *Oxidation mechanism of an Fe–9Cr–1Mo steel by liquid Pb–Bi eutectic alloy at 470 C (Part II)*. Corrosion Science, 2008. **50**(9): p. 2537-2548.
36. Ellingham, H.J., *Reducibility of oxides and sulphides in metallurgical processes*. J. Soc. Chem. Ind, 1944. **63**(5): p. 125-160.

37. Young, D.J., *High temperature oxidation and corrosion of metals*. Vol. 1. 2008: Elsevier.
38. Birks, N., G.H. Meier, and F.S. Pettit, *Introduction to the high temperature oxidation of metals*. 2006: Cambridge university press.
39. Tan, L., X. Ren, and T.R. Allen, *Corrosion behavior of 9–12% Cr ferritic–martensitic steels in supercritical water.* Corrosion science, 2010. **52**(4): p. 1520-1528.
40. Bischoff, J. and A.T. Motta, *Oxidation behavior of ferritic–martensitic and ODS steels in supercritical water.* Journal of Nuclear Materials, 2012. **424**(1-3): p. 261-276.
41. Ye, Z., et al., *Oxidation mechanism of T91 steel in liquid lead-bismuth eutectic: with consideration of internal oxidation.* Scientific reports, 2016. **6**(1): p. 35268.
42. Tan, L., Y. Yang, and T. Allen, *Oxidation behavior of iron-based alloy HCM12A exposed in supercritical water.* Corrosion Science, 2006. **48**(10): p. 3123-3138.
43. Chen, Y., K. Sridharan, and T. Allen, *Corrosion behavior of ferritic–martensitic steel T91 in supercritical water.* Corrosion Science, 2006. **48**(9): p. 2843-2854.
44. Sun, L. and W. Yan, *Estimation of Oxidation Kinetics and Oxide Scale Void Position of Ferritic-Martensitic Steels in Supercritical Water.* Advances in Materials Science and Engineering, 2017. **2017**(1): p. 9154934.
45. Li, Y., et al., *Predictions and analyses on the growth behavior of oxide scales formed on ferritic–martensitic in supercritical water.* Oxidation of Metals, 2019. **92**(1): p. 27-48.
46. Agüero, A., et al., *Oxidation under pure steam: Cr based protective oxides and coatings.* Surface and coatings technology, 2013. **237**: p. 30-38.
47. Ma, L., et al., *Effects of Cr content on the microstructure and properties of 26Cr–3.5 Mo–2Ni and 29Cr–3.5 Mo–2Ni super ferritic stainless steels.* Journal of Materials Science & Technology, 2016. **32**(6): p. 552-560.
48. Gilbert, M., et al., *Vacancy defects in Fe: Comparison between simulation and experiment.* Journal of nuclear materials, 2009. **386**: p. 36-40.
49. Krishnamurthy, R. and D. Srolovitz, *Stress distributions in growing oxide films.* Acta materialia, 2003. **51**(8): p. 2171-2190.
50. Galerie, A., et al., *Stress and adhesion of chromia-rich scales on ferritic stainless steels in relation with spallation.* Materials Research, 2004. **7**: p. 81-88.
51. Bamba, G., et al., *Thermal oxidation kinetics and oxide scale adhesion of Fe–15Cr alloys as a function of their silicon content.* Acta Materialia, 2006. **54**(15): p. 3917-3922.
52. Vogt, J.-B. and I. Proriol Serre, *A review of the surface modifications for corrosion mitigation of steels in lead and LBE.* Coatings, 2021. **11**(1): p. 53.
53. Wang, H., et al., *Corrosion behavior and surface treatment of cladding materials used in high-temperature lead-bismuth eutectic alloy: A review.* Coatings, 2021. **11**(3): p. 364.

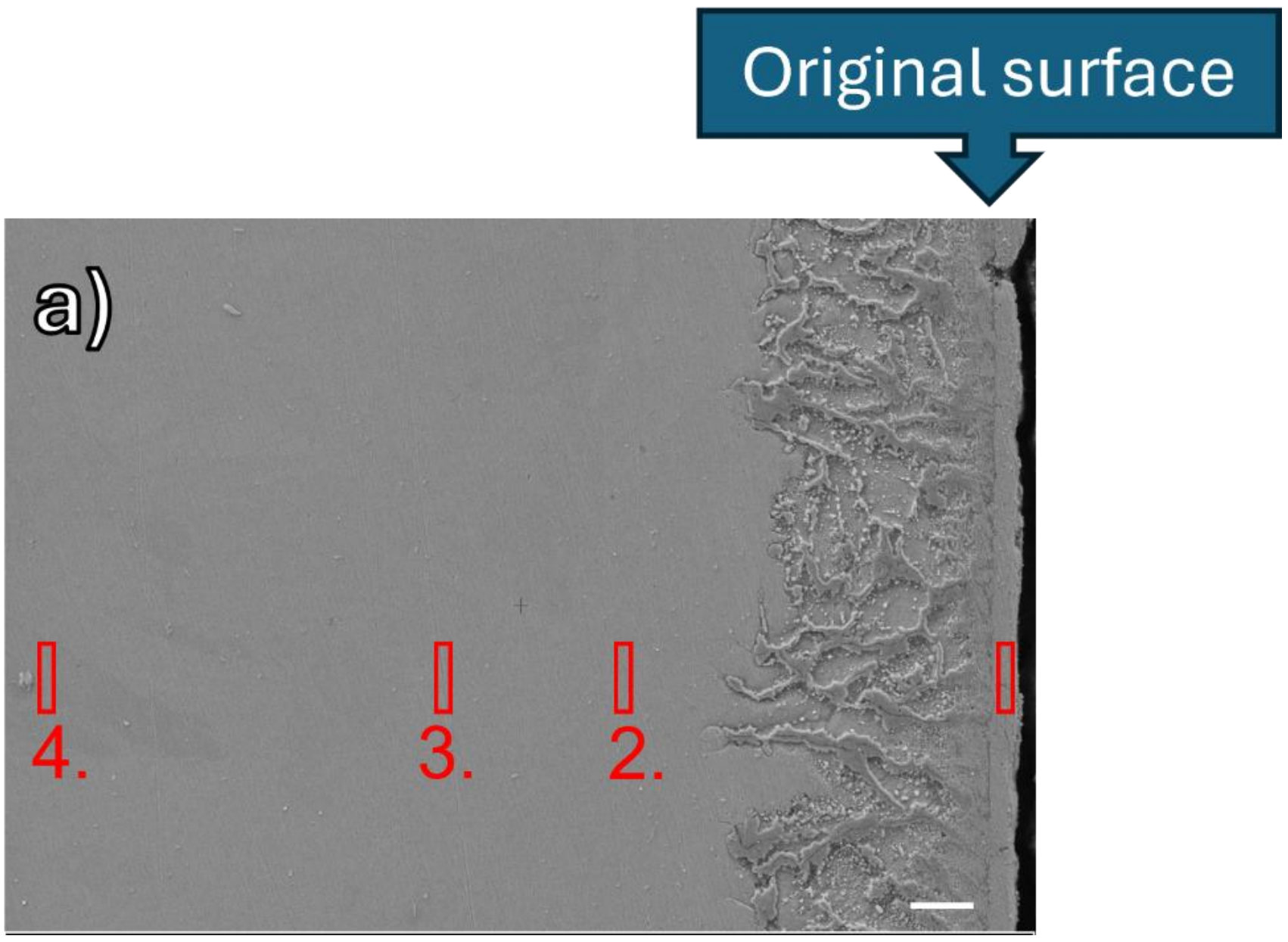

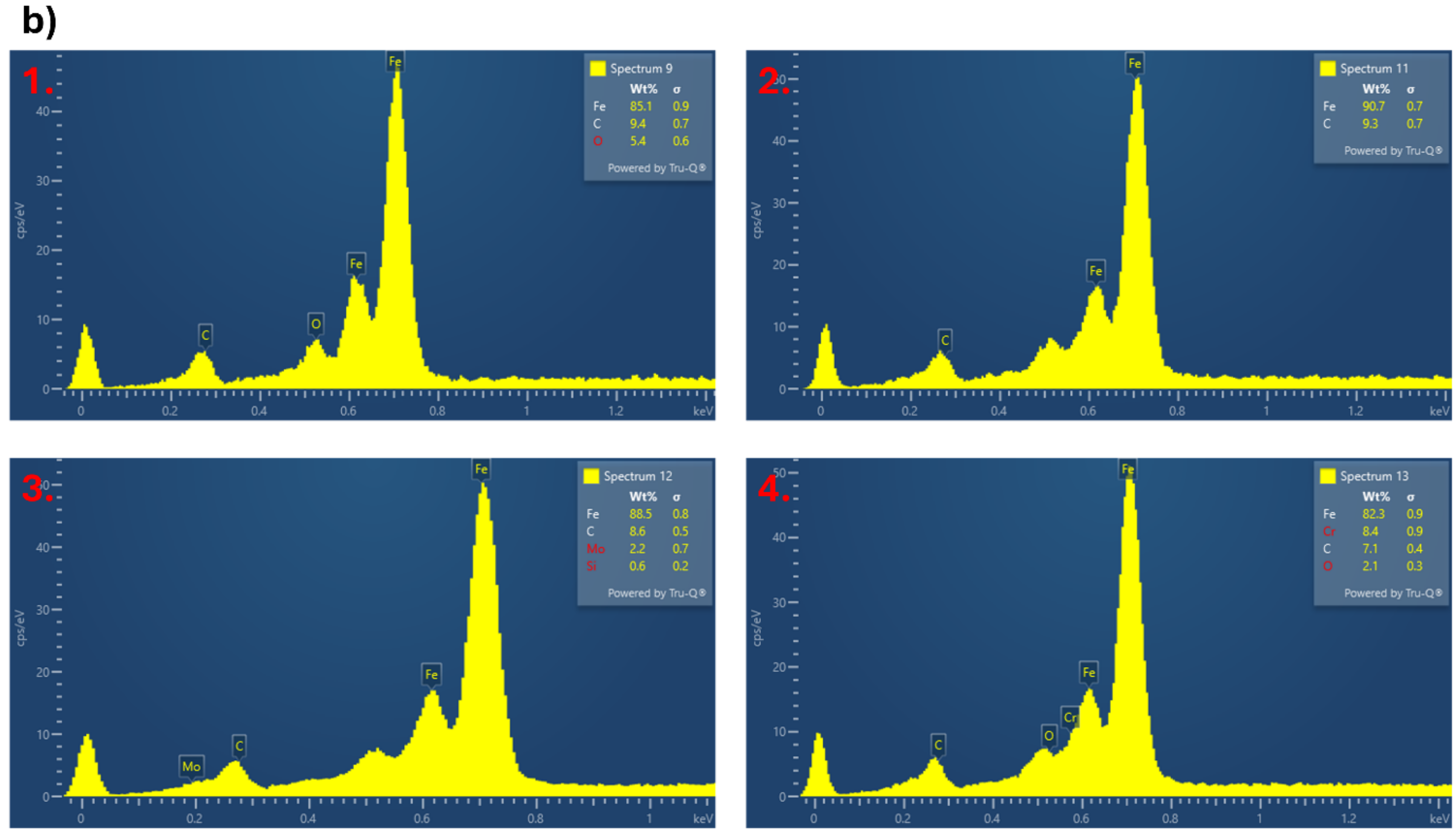

*S1. (a) SEM view with the place where spectra were taken highlighted with red rectangles.*

*(b) The spectra for 4 areas marked in S1 (a).*

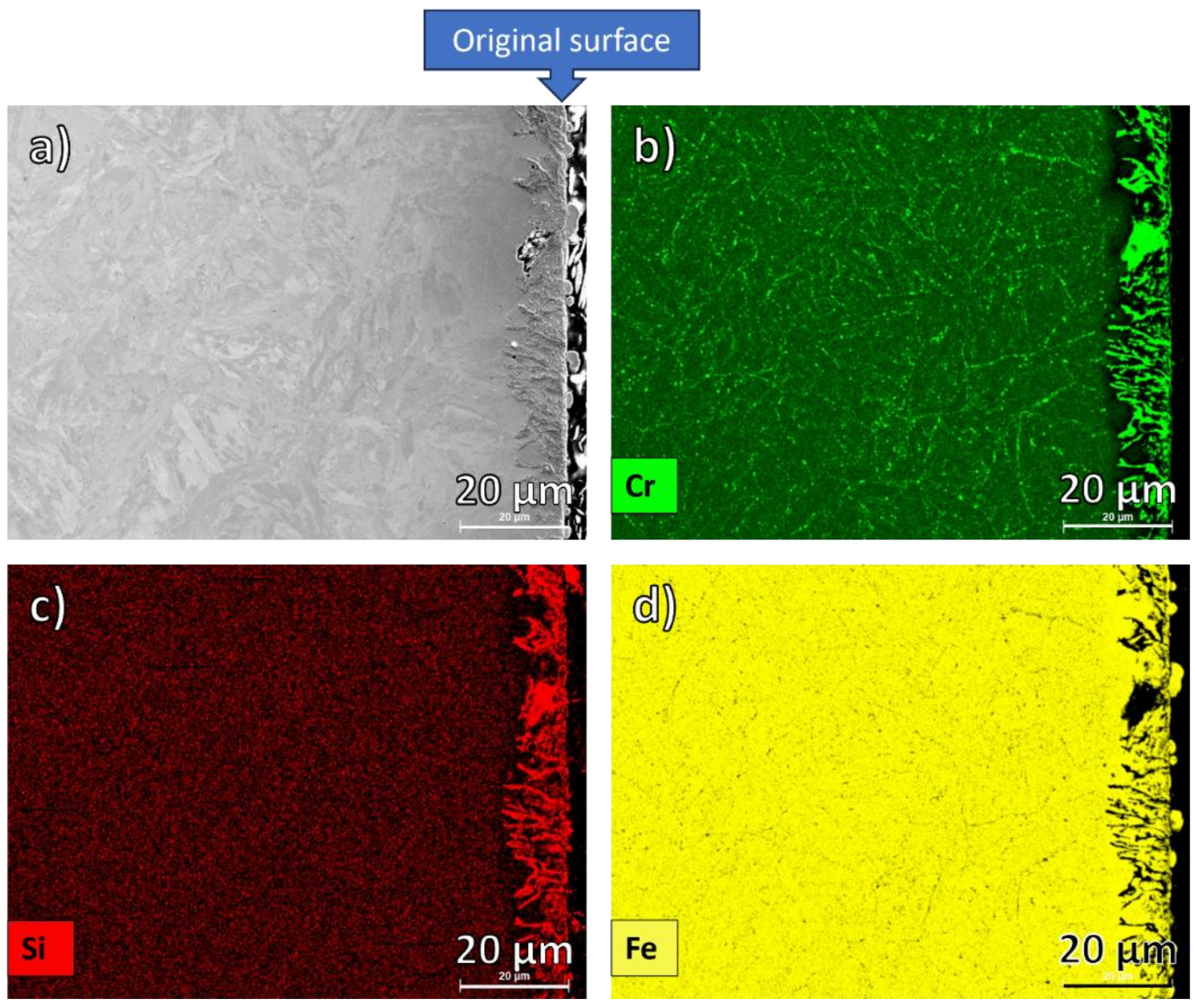

*S2. SEM-EDX results [Sample: 70 h, oxidising environment, 700 ℃, LBE]. (a) SEM image. (b), (c), (d) EDX results highlighting Cr, Si, and Fe.*

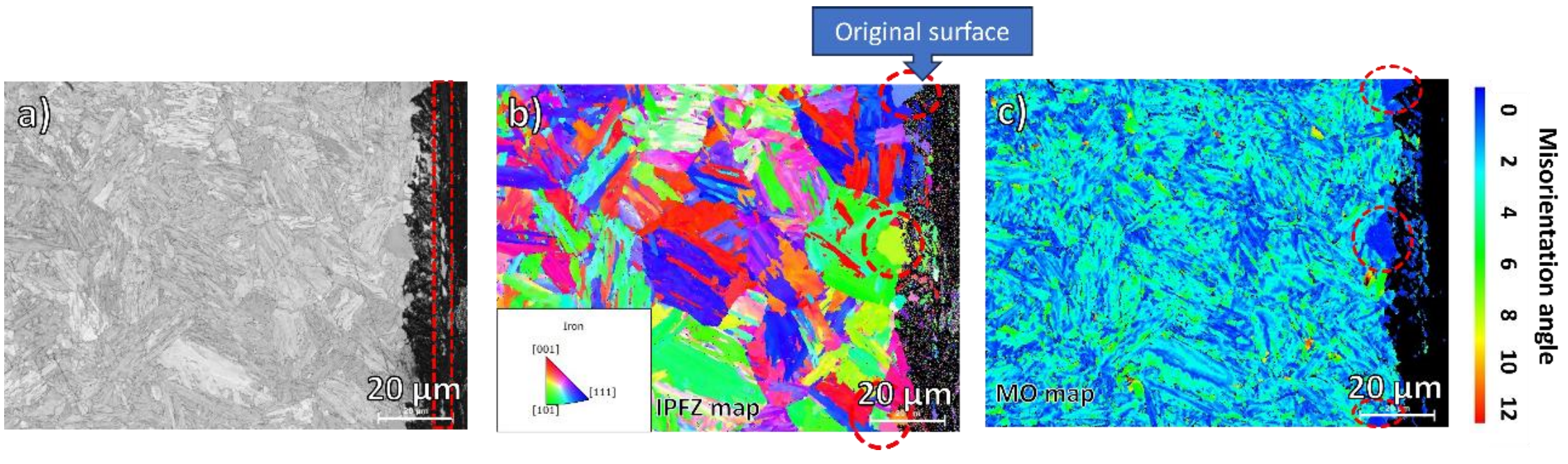

*S3. SEM-EBSD correlated with SEM-EDX shown in S1 [Sample: 70 h, oxidising environment, 700 ℃, LBE]. (a) EBSD pattern quality map with red dashed rectangular highlighting the layer formed outside the surface. (b) IPFZ map with red dashed circles highlighting the grains where*

*phase change may have happened. (c) Grain average misorientation map with a scale bar on the right. The red dashed circles highlight the same grains as in (b).*

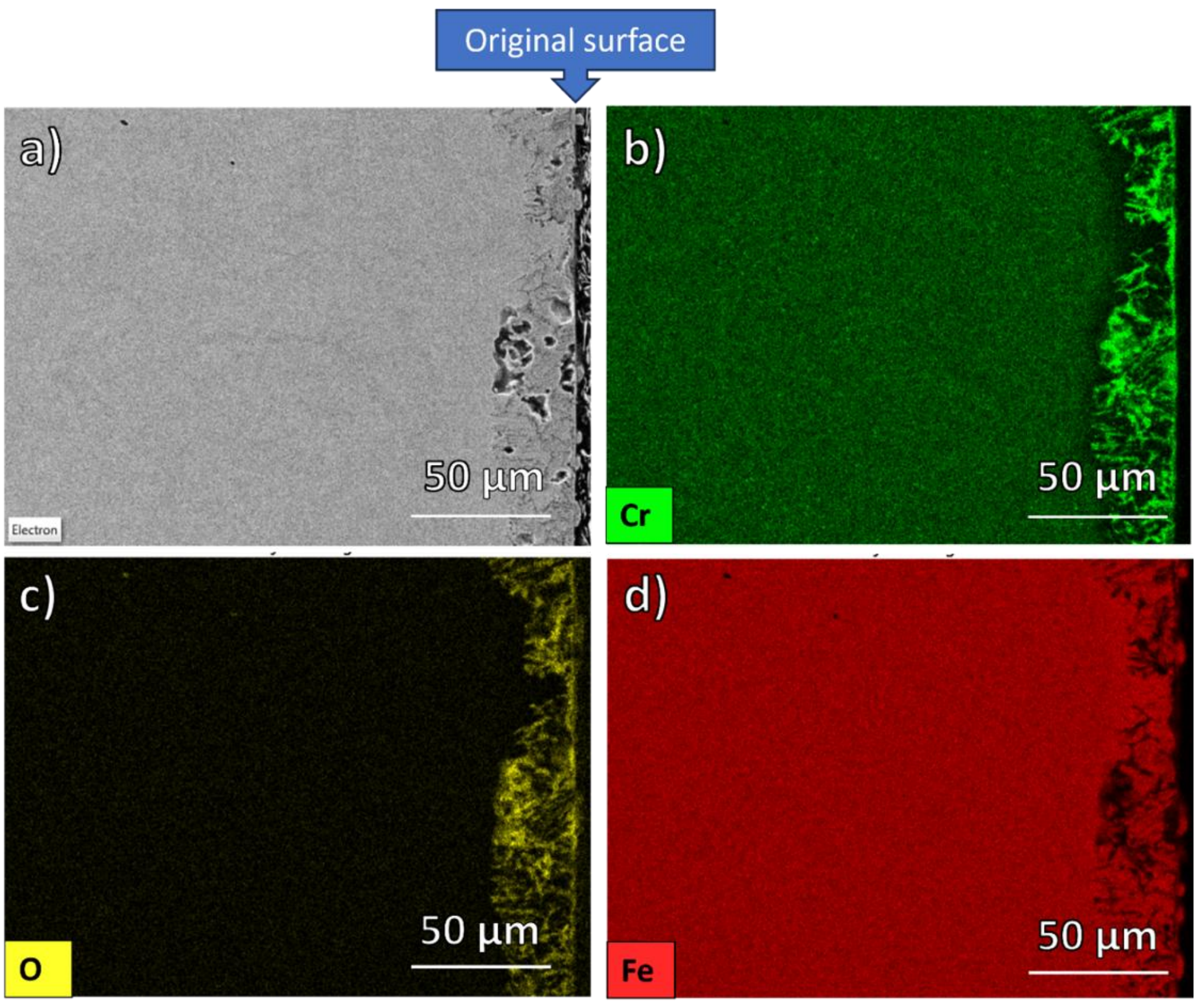

*S4. SEM-EDX results [Sample: 506 h, oxidising environment, 700 ℃, LBE]. (a) SEM image. (b), (c), (d) EDX results highlighting Cr, Si, and Fe.*

To quantify elemental distributions in the corroded GBs of the 245 h corroded sample, an EDX line scan was conducted through the GBs as shown in *S5*(a). *S5*(b) shows the corresponding changes in elemental concentrations of Cr, Fe, Mo, Si, and O. From the free surface to the unaffected T91 matrix, an Fe oxide, Fe-Cr oxide, and Si oxide can be observed. By analysing changes in the Cr content, a Cr depletion zone can be seen near the Cr-enriched oxides. There are also several Mo-enriched precipitates near the corroded GBs. As the Pb-M x-ray energy is 2.342 keV and the Bi-M x-ray energy is 2.419 keV, similar to the Mo-$L_{\alpha}$ energy of 2.293 keV, peak overlap is a concern. A higher voltage of 30 keV was thus also used for EDX, at the expense of some spatial resolution, and the result shows that the precipitates can be confirmed to be Mo-enriched (*Supplementary Materials S6*).

In *S5*, Si-enriched and Cr-enriched oxides appear in different places, suggesting the formation of a layered oxide in the GBs. This is similar to our observation in the 506 h corroded sample shown in Fig. 4(e).

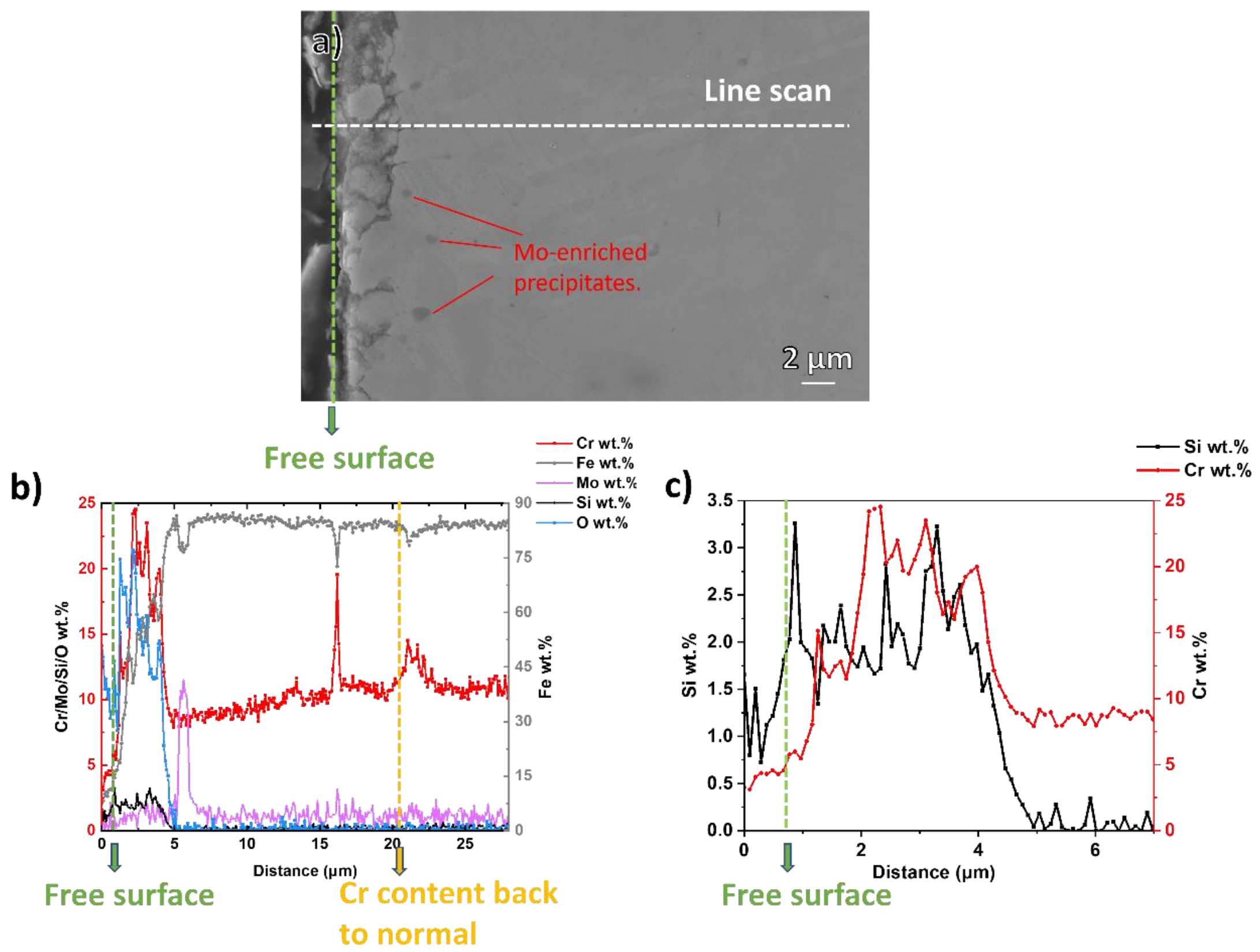

*S5. SEM-EDX line scan results for 245 h corroded sample in oxidising environment. (a) SEM SE image showing the position of the line scan and Mo-enriched precipitates. (b) EDX line scan result showing content of Cr, Fe, Mo, Si, and O. (c) EDX line scan showing Si and Cr content at the surface region.*

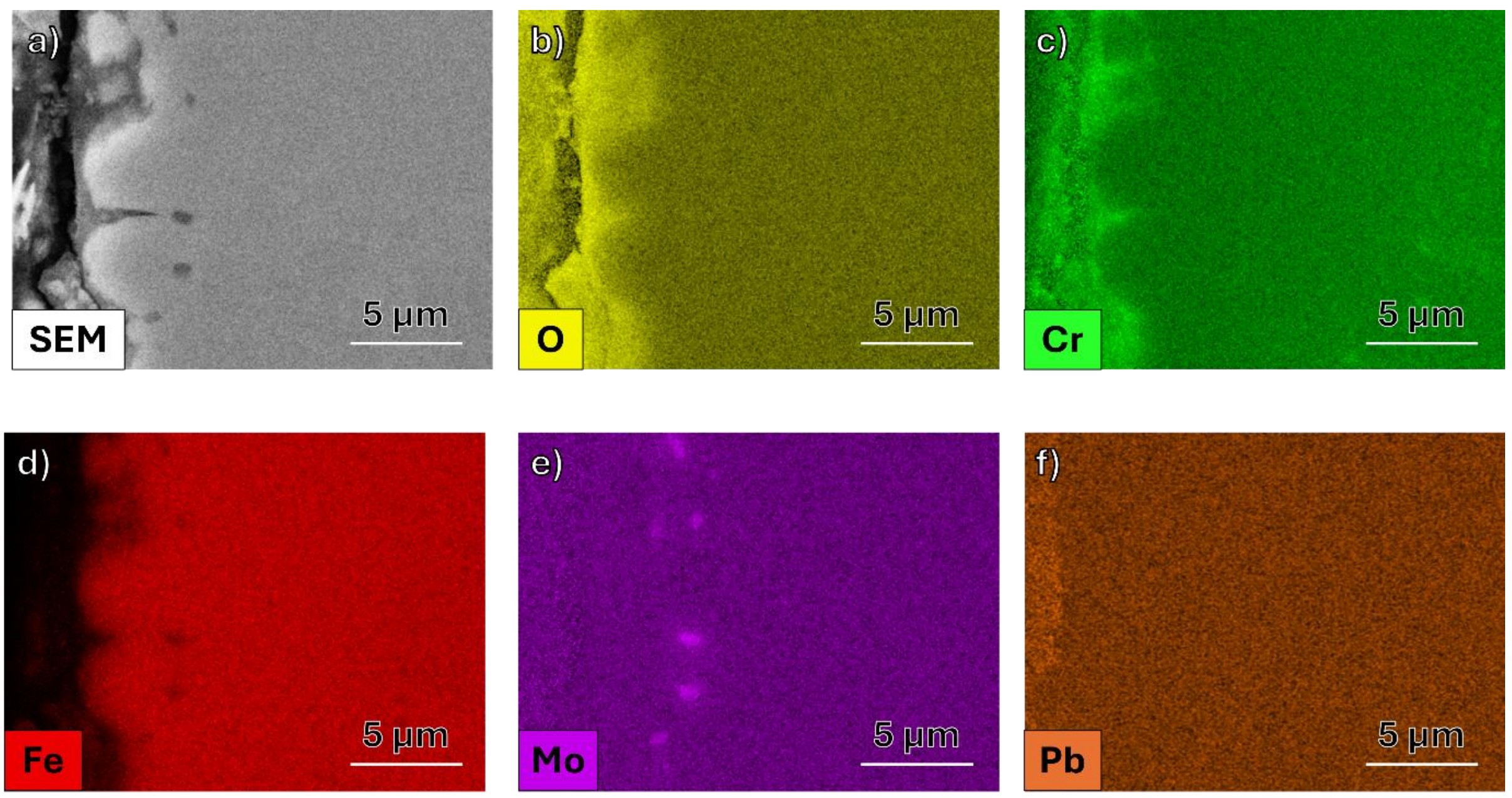

*S6. 30 kV EDX maps [Sample: 245 h, oxidising environment, 700 °C, LBE]. (a) SEM image. (b) – (f) EDX results highlight O, Cr, Fe, Mo, and Pb, respectively.*

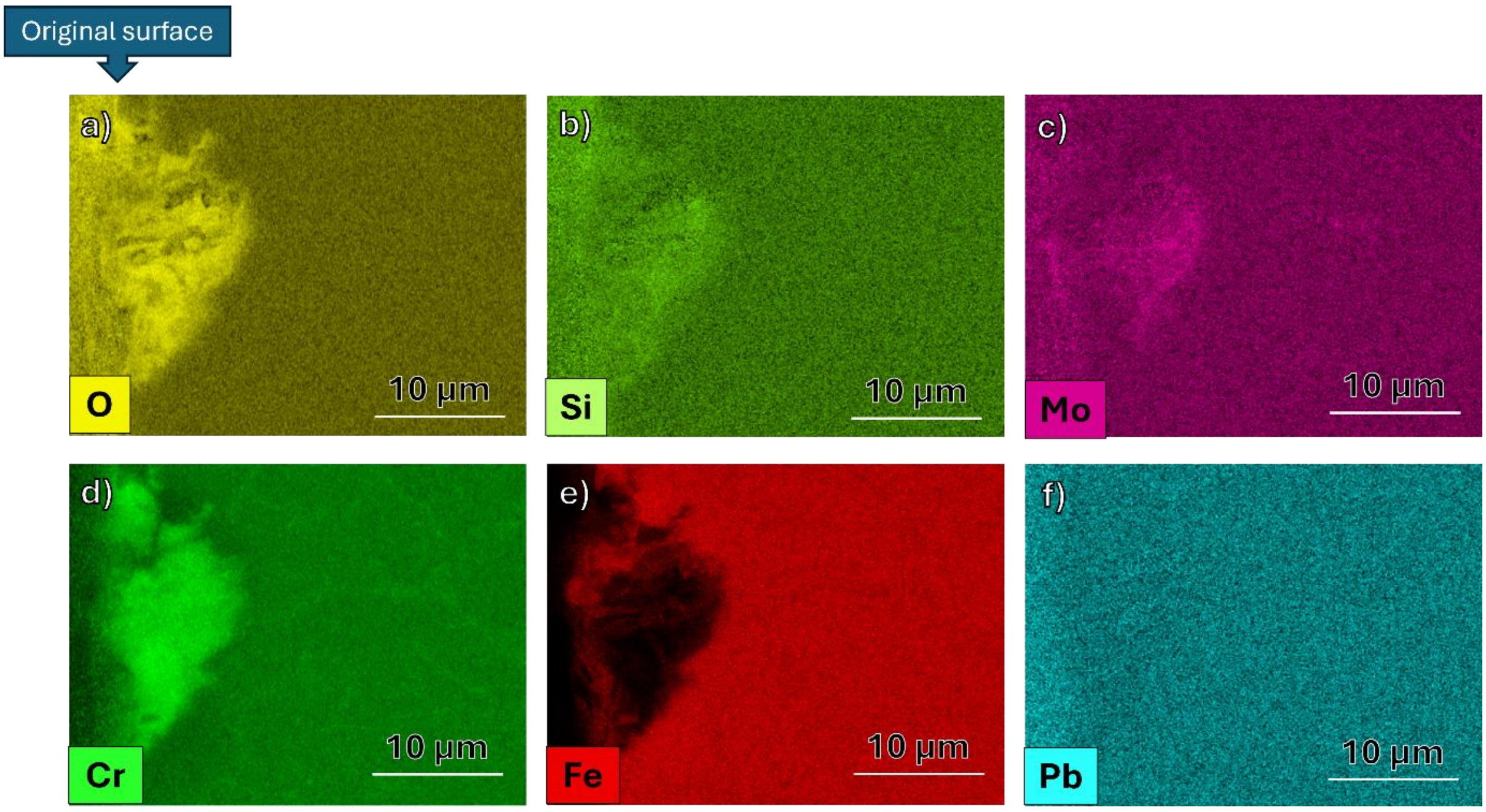

*S7. SEM-EDX results acquired at 25 kV of area corrosion [Sample: 70 h, oxidising environment, 700 ℃, LBE]. (a) -(f) SEM-EDX results (25 kV) for Fig. 7.1 (d) highlighting O, Si, Mo, Cr, Fe and Pb, respectively.*

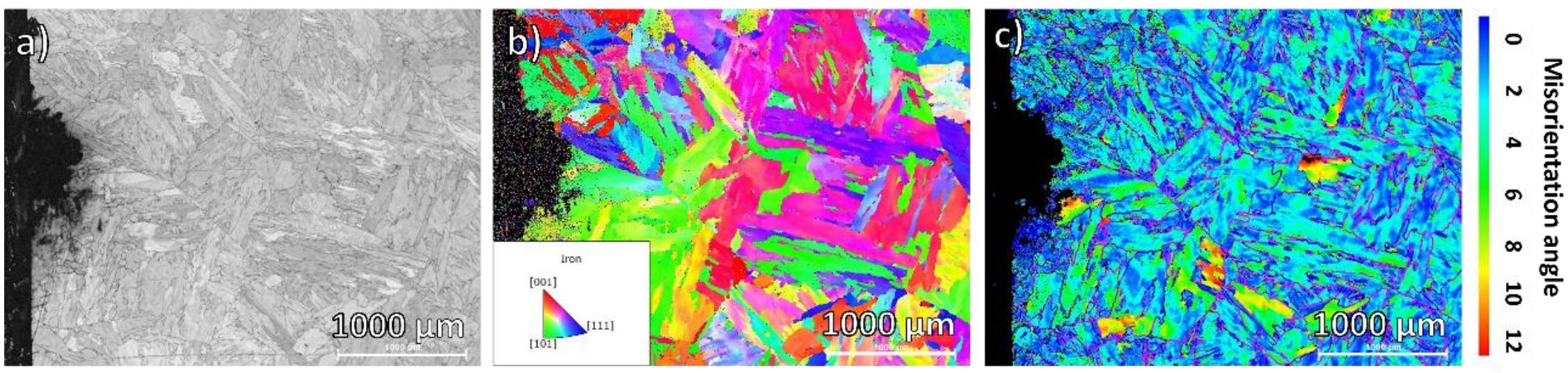

*S8. SEM-EBSD results for Fig. 7.25 from a lower magnification [Sample: 70 h, oxidising environment, 700 ℃, LBE]. (a) EBSD pattern quality map. (b) EBSD-IPFZ map. (c) EBSD grain average misorientation map with the scale bar.*

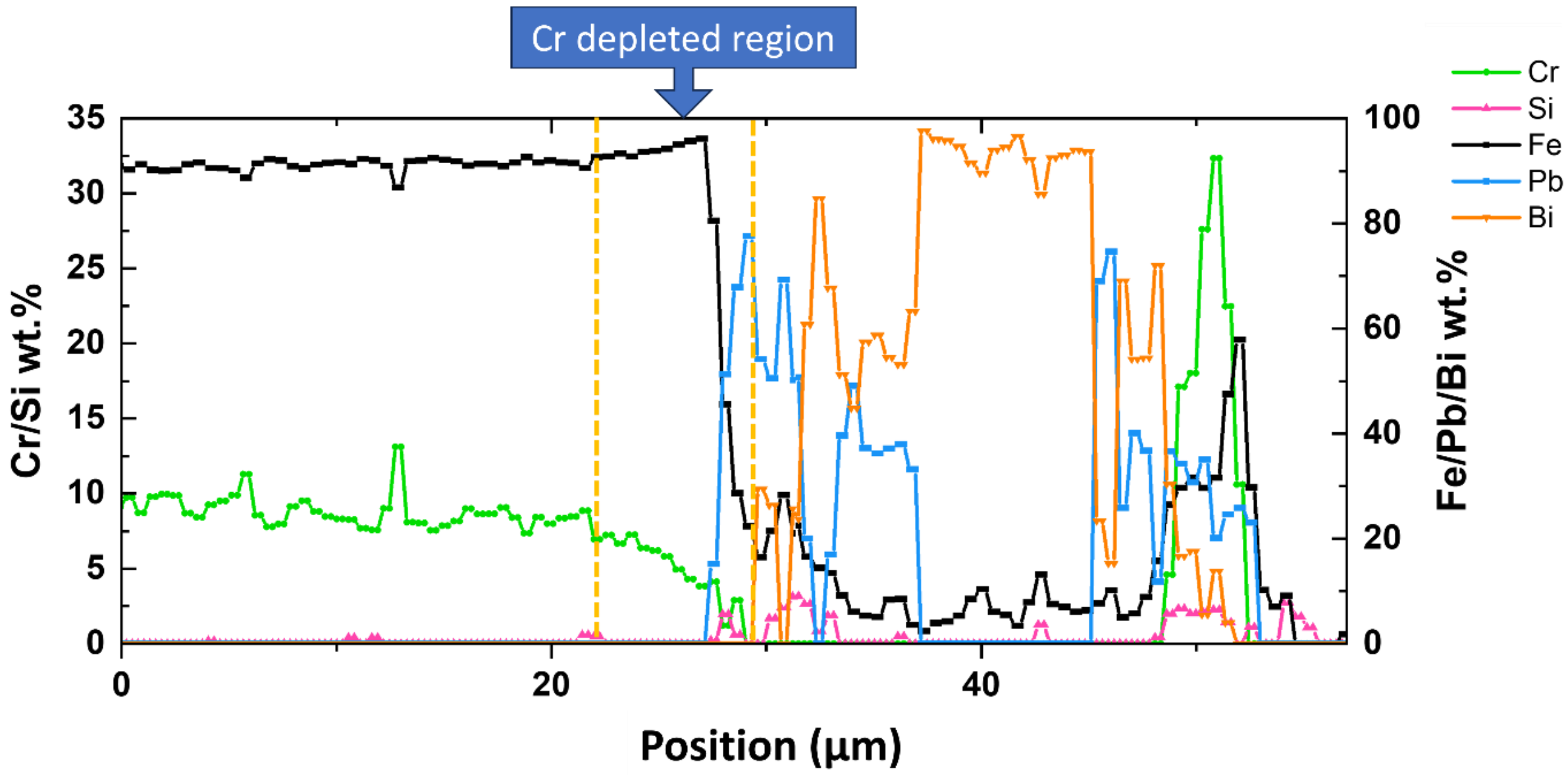

*S9. The line-scan along the white line shown in Fig. 8(a).*

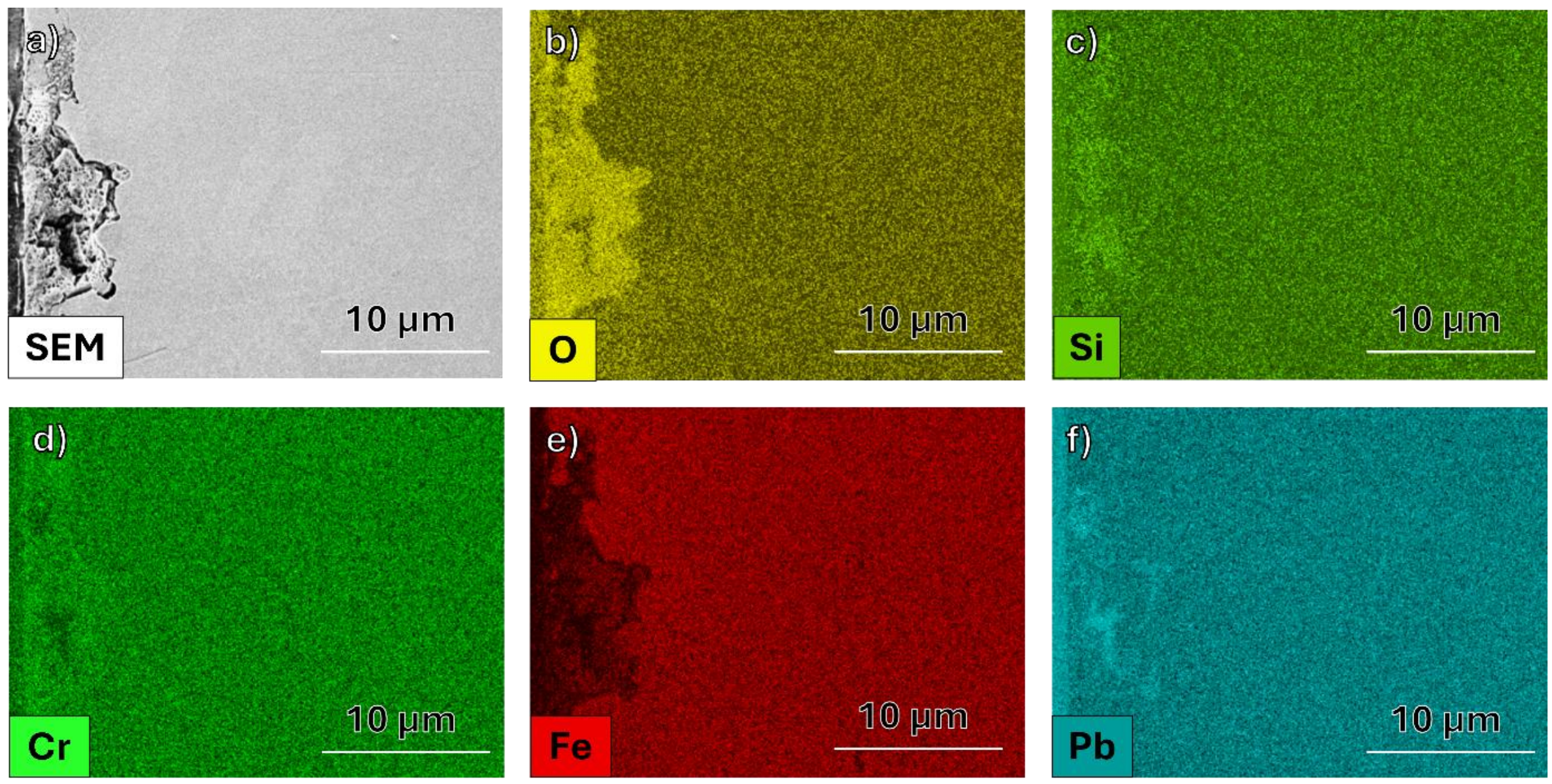

*S10. SEM-EDX results acquired at 5 kV with area corrosion [Sample: 245 h, oxidising environment, 700 ℃, LBE]. (a) SEM image. (b)-(f) SEM-EDX results highlighting O, Si, Cr, Fe and Pb, respectively. (S10 (f) can be the peak of Pb, Bi, or Mo, 5 KV is unable to distinguish).*

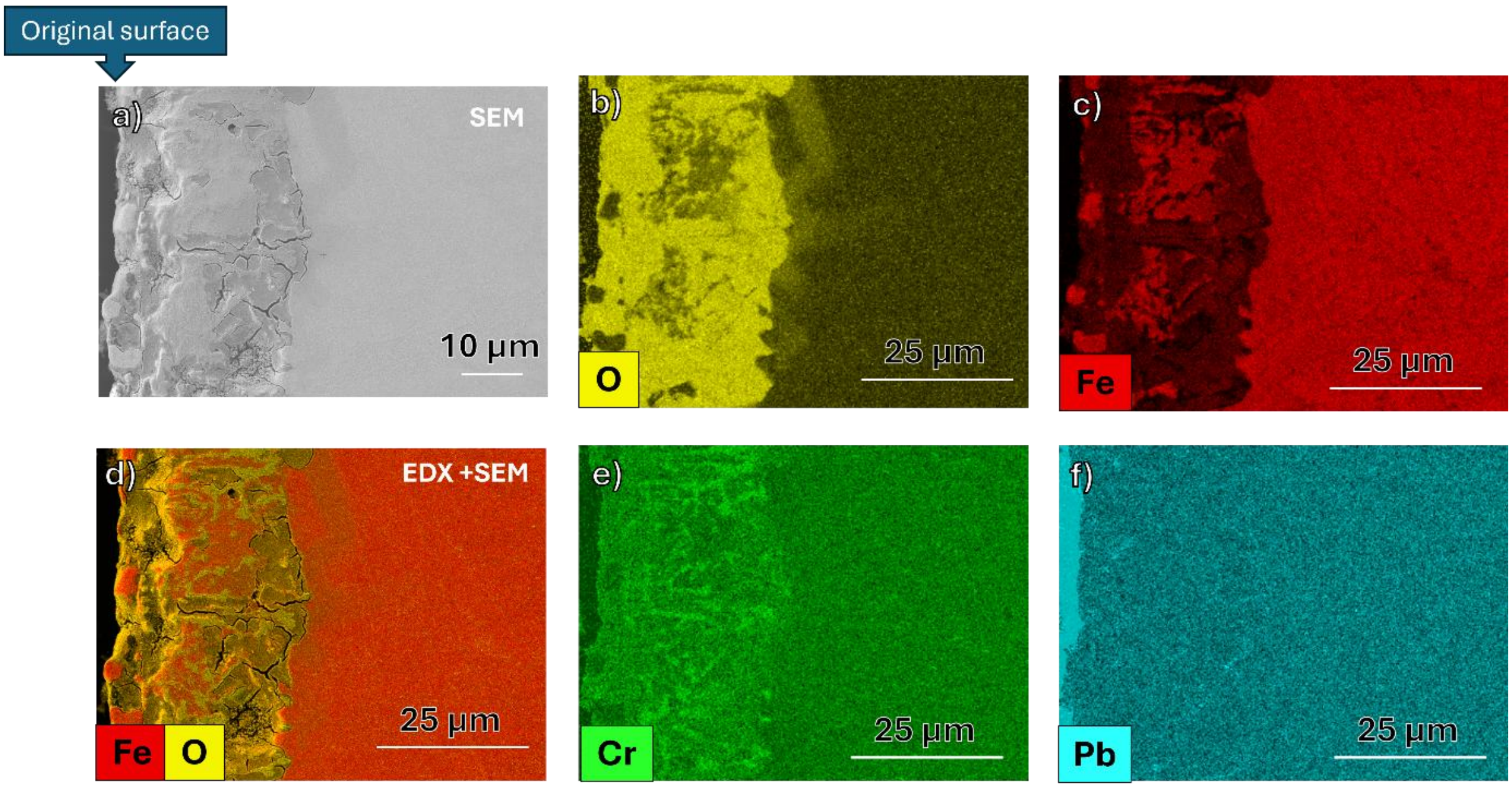

*S11. SEM-EDX with cracks observed in the oxide layer [Sample: 506 h, oxidising environment, 700 ℃, LBE]. (a) SEM image. (b)(c)(e)(f) EDX results highlight O, Fe, Cr, and Pb, respectively. (d) EDX results highlight Fe and Cr overlay with the SEM image.*

*S12* shows the SEM-EDX result for the 70 h corroded sample. There is no visible oxidation or corrosion in *S12* (a). *S12* (b)-(d) confirm the formation of a dense layer of Si and Cr enriched oxide. *S12* (d) shows a Cr depleted region adjacent to the oxide layer which may have formed because Cr diffused to the surface to form the oxide layer. *S12* (e) shows that there is negligible Fe in the protective oxide layer. Finally, *S12* (f) shows the signal for Pb which does not penetrate the material.

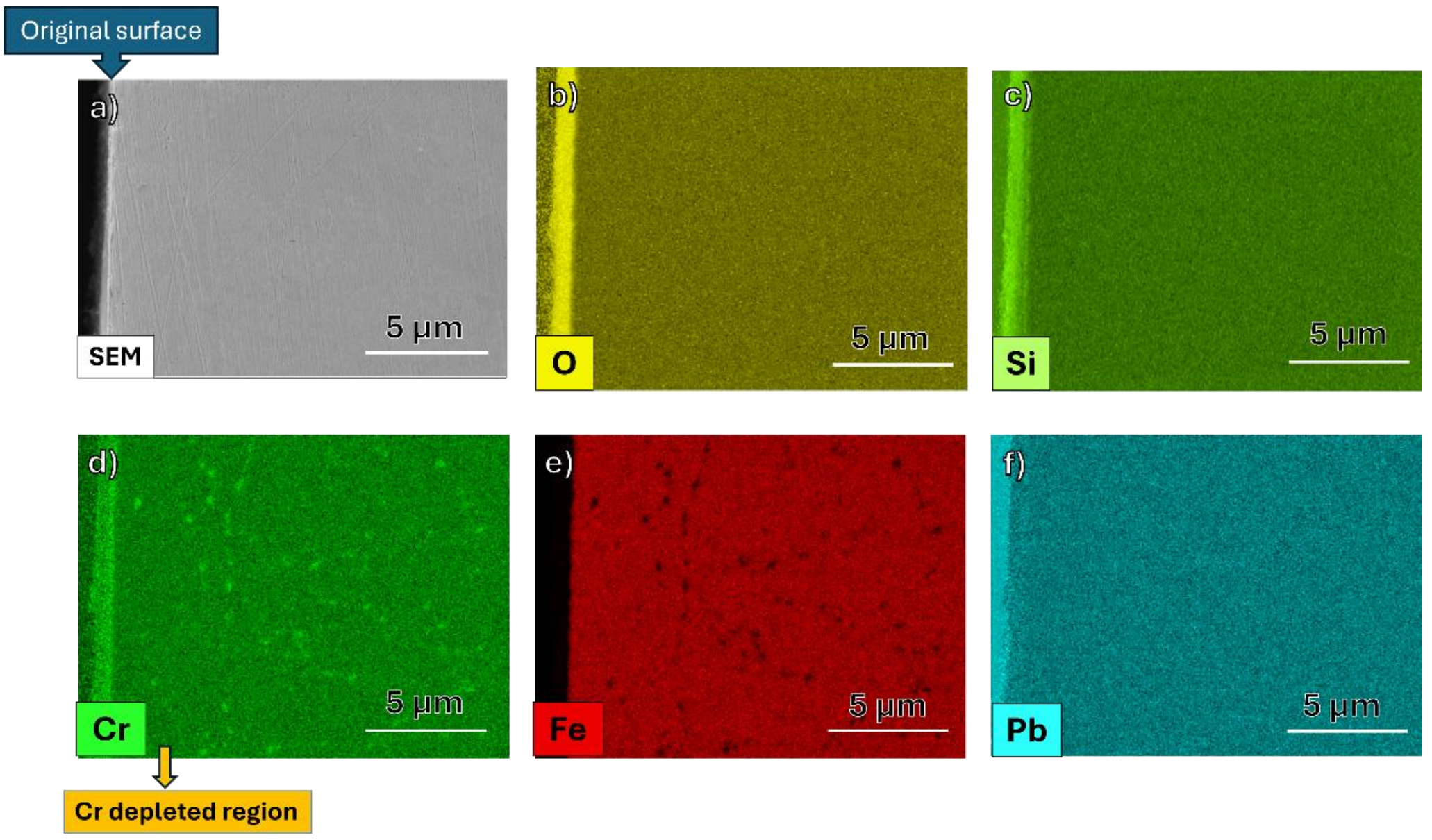

*S12. The SEM-EDX results acquired at 5 KV [Sample: 70 h, oxidising environment, 700 ℃, LBE]. (a) SEM view. (b)-(f) SEM-EDX results highlighting O, Si, Cr, Fe, and Pb, respectively.*

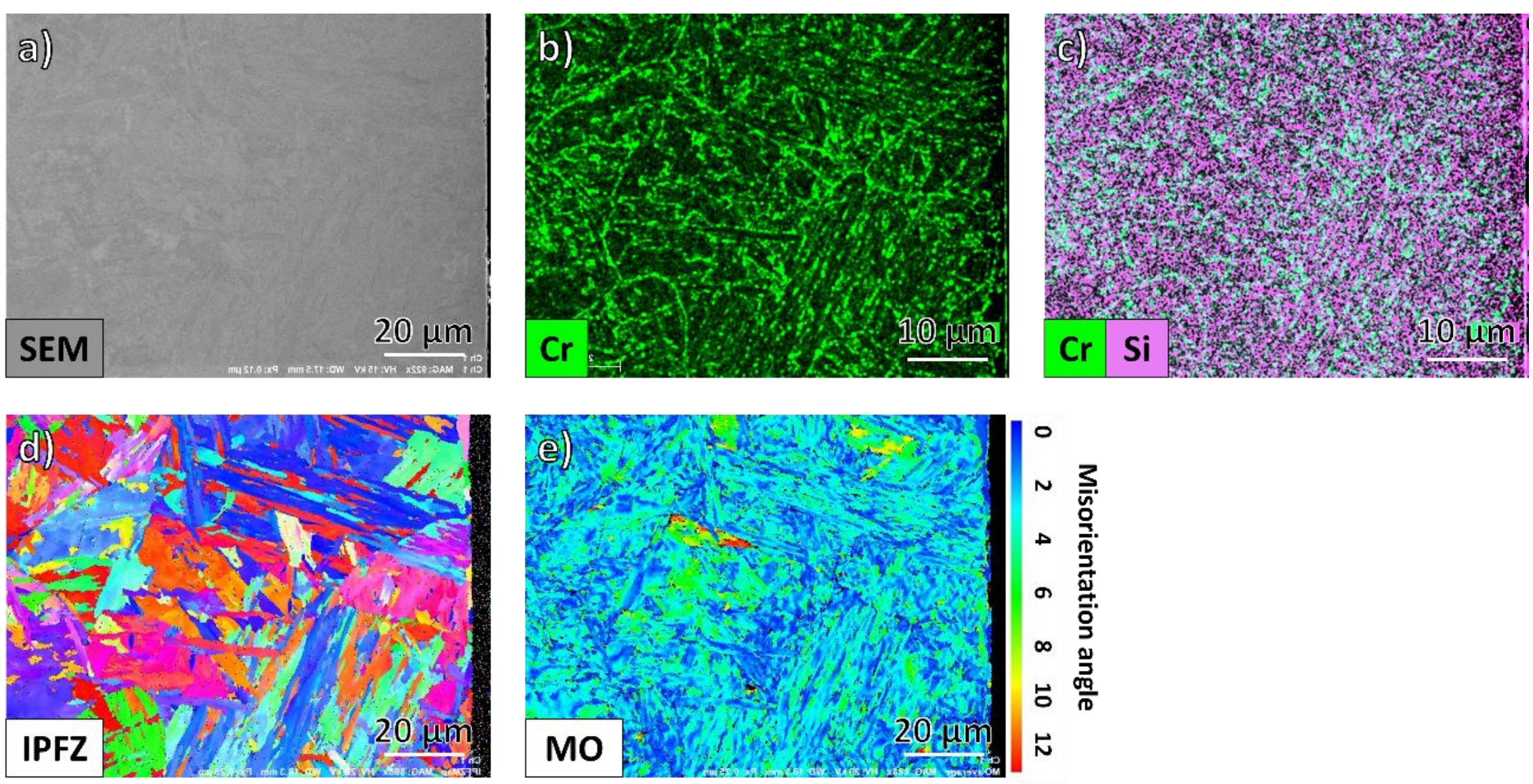

*S13. SEM-EDX and EBSD results for one representative area with no obvious corrosion in SEM view [Sample: 506 h, oxidising environment, 700 ℃, LBE]. (a) SEM view. (b) SEM-EDX result highlighting Cr. (c) SEM-EDX result highlighting Cr and Si. (d) SEM-EBSD IPFZ map. (e) SEM-EBSD grain average misorientation with scale bar.*